\documentclass[a4paper,12pt]{article}
\usepackage{amsmath,amsfonts,amssymb}
\usepackage{graphicx}

\def\unit{\relax{\rm 1\kern-.26em I}}

\usepackage{verbatim,setspace}
\usepackage{xparse}
\usepackage{tikz}
\usetikzlibrary{calc}
\DeclareDocumentCommand{\hcancel}{mO{0pt}O{0pt}O{0pt}O{0pt}}{%
    \tikz[baseline=(tocancel.base)]{
        \node[inner sep=0pt,outer sep=0pt] (tocancel) {#1};
        \draw[black] ($(tocancel.south west)+(#2,#3)$) -- ($(tocancel.north east)+(#4,#5)$);
    }
}

\usepackage{amsmath}
\usepackage{amssymb}
\usepackage{amsfonts}
\usepackage{epsfig}
\usepackage{soul}
\newcommand{\cA}{{\cal A}}

\newcommand{\cC}{{\cal C}}
\newcommand{\cD}{{\cal D}}

\newcommand{\cF}{{\cal F}}

\newcommand{\cL}{{\cal L}}

\newcommand{\cO}{{\cal O}}

\newcommand{\cS}{{\cal S}}
\newcommand{\cT}{{\cal T}}

\newcommand{\cZ}{{\cal Z}}

\newcommand{\ov}{\overline}

\newcommand{\Tr}{\mbox{Tr}}

\newcommand{\be}{\begin{equation}}
\newcommand{\ee}{\end{equation}}
\newcommand{\bea}{\begin{eqnarray}}
\newcommand{\eea}{\end{eqnarray}}

\DeclareMathSymbol{\mg}{\mathrel}{symbols}{"1D}

\newcommand{\MET}{E\llap{/\kern1.5pt}_T}
\newcommand{\nn}{\nonumber}

\newcommand{\msb}{m_{\hcancel{\tiny{SUSY}}}}

\newcounter{oldcounter}
\addtocounter{equation}{1}
\begin{document}

 \begin{flushright}
{CERN-PH-TH/2012-317}\\
CPHT-RR086.1112
\end{flushright}

\thispagestyle{empty}

\begin{center}
\begin{spacing}{1}
{\LARGE {\bf Low scale supersymmetry breaking and its\\ LHC signatures}}
\end{spacing}
\vspace{1cm}
{ {\bf Emilian Dudas$^{\,a,b}$,
Christoffer Petersson$^{\,c,d}$ and  
Pantelis Tziveloglou$^{\,a}$}}
\end{center}
{\small \noindent{$^a$ Centre de Physique Th\'eorique, \'Ecole Polytechnique, CNRS, 91128 Palaiseau, France}
\\[2pt]
{$^b$ Theory Division, CERN, 1211 Geneva 23, Switzerland}
\\[2pt]
{$^c $ Physique Th\'eorique et Math\'ematique, Universit\'e Libre de Bruxelles, \\
\phantom{v} C.P. 231, 1050 Bruxelles, Belgium}
\\[2pt]
{$^d$ International Solvay Institutes, Brussels, Belgium}}
\\[5pt]
{\footnotesize E-mail addresses:\,\,emilian.dudas@cpht.polytechnique.fr, christoffer.petersson@ulb.ac.be, \\pantelis.tziveloglou@cpht.polytechnique.fr}

 \begin{abstract}
\noindent
We study the most general extension of the MSSM Lagrangian that includes scenarios in which supersymmetry is spontaneously broken at a low scale $f$. The spurion that parametrizes supersymmetry breaking in the MSSM is promoted to a dynamical superfield involving the goldstino, with (and without) its scalar superpartner, the sgoldstino. The low energy effective Lagrangian is written as an expansion in terms of $\msb/\sqrt{f}$, where $\msb$ is the induced supersymmetry breaking scale, and contains, in addition to the usual MSSM Lagrangian with the soft terms, couplings involving the component fields of the goldstino superfield and the MSSM fields. This Lagrangian can provide significant corrections to the usual couplings in the Standard Model and the MSSM. We study how these new corrections affect the Higgs couplings to  gauge bosons and fermions, and how LHC bounds can be used in order to constrain $f$. We also discuss that, from the effective field theory point of view, the couplings of the goldstino interactions are not determined by any symmetry, and their usual simple relation to the soft terms is corrected by higher-dimensional operators.  
\end{abstract}

\newpage

\tableofcontents

\setcounter{page}{1}

\newpage

\section{Introduction}

In the usual formulation of the low energy minimal supersymmetric Standard Model (MSSM), the supersymmetric Lagrangian is supplemented with soft-breaking terms \cite{gg}, with parameters  generically denoted by $m_{soft}$ in what follows\footnote{We also include the $\mu$-term in
$m_{soft}$, since numerically its value cannot be very different from the soft terms.}. 
In supergravity with supersymmetry (SUSY) breaking, the relation between the SUSY breaking scale $f$ and the Planck mass $M_P$  depends on how the SUSY breaking is mediated to the
observable sector: in general, $m_{soft} \sim f/M_P$ for gravity mediation while $m_{soft} \sim (g^2/16 \pi^2) f/M $ for gauge mediation,
where  $g^2/16 \pi^2$ is a SM loop-factor and $M$ is the intermediate mass scale of some messenger fields. One can take the low-energy limit by
sending $M_P \to \infty$, while keeping fixed $m_{soft}$ \cite{bfs}. For gravity mediation, this implies that $\sqrt{f} \sim 10^{11}$ GeV, whereas for gauge mediation, by fixing the soft terms in the TeV range and requiring the absence of tachyons in the messenger sector, one gets that $\sqrt{f} \gtrsim 50-100 \ {\rm TeV} $ \cite{gr}. 

In the rigid limit, obtained by decoupling supergravity interactions, the couplings to the transverse components of the spin 3/2 gravitino vanish, whereas its longitudinal component, the goldstino $G$, has couplings given
by the equivalence theorem \cite{cddfg}, $\frac{1}{f} \partial_{\mu} G \, J^{\mu}$, where $J^{\mu}$ is the supercurrent.  Hence, the smaller the scale $f$ (or, alternatively, the lighter
the gravitino, since they are related via $f \sim m_{3/2} M_P$), the stronger the goldstino couplings to matter.

Not much is known about mediation schemes which are phenomenologically viable in the case where ${\sqrt f}$ is below 10 TeV.  Nevertheless, it is interesting to contemplate what characteristic phenomenological features such a scenario could have. Apart from the discussion in, for example, \cite{Gherghetta:2011na}, the possibility of realizing this scenario in terms of
 strongly coupled mediation models is to a large extent unexplored and is certainly worth a special dedicated effort. We will not have 
anything new to say here about explicit SUSY breaking models in this class; our approach is to parameterize the resulting
low-energy effective action in a model-independent way.

In the present paper, we extend earlier work on phenomenological consequences of low-scale supersymmetry breaking \cite{Espinoza,SK,gero,Antoniadis:2010hs,sgold,pr,Bellazzini:2012mh,Petersson:2012nv} by considering the most general set of operators contributing to the low-energy goldstino couplings to matter, in an expansion in $\msb/{\sqrt f}$, up to dimension-six operators, where $\msb$ denotes the induced SUSY breaking scale in the visible sector. The resulting operators can be organized according to the number of goldstinos
they contain. We will be interested in couplings with no goldstinos and with one goldstino. 

The couplings without goldstinos correspond to genuinely new contributions to the interactions
among the MSSM fields themselves, arising from the dynamics of goldstino multiplet $X$, e.g.~when the corresponding auxiliary field $F_X$ and the sgoldstino (in the case where the sgoldstino is present in the effective action) are integrated out at tree level. 
We show that these new couplings can significantly affect, for instance, Higgs couplings to gauge bosons and fermions compared to the MSSM, if ${\sqrt f}$ is of the order of a few TeV.  In addition, four-fermion contact interactions are generated and they are constrained by flavor physics
as well as by direct searches at the LHC and Tevatron.
   
The one-goldstino couplings to the MSSM fields we find contain the usual supercurrent couplings  
$\frac{1}{f} \partial_{\mu} G \, J^{\mu}$. This is obtained in the minimal setup containing MSSM plus the minimal set of operators needed to parameterize the soft-breaking terms, called non-linear MSSM in \cite{Antoniadis:2010hs}. By using equations of motion and field redefinitions, we show that the effect of additional higher-dimensional/derivative operators is to correct existing couplings $\lambda$ in the following generic way,
\be
\lambda \ = \ \lambda_{MSSM} \left( 1 + \sum_n c_n \left(\frac{\msb}{\sqrt{f}}\right)^n \right) \ . \label{intro1}
\ee
In the case where the effective operators are generated by integrating out a SUSY messenger sector, with a characteristic scale significantly higher than the SUSY breaking scale, 
only even powers appear in (\ref{intro1}). If we do not make any assumptions about the origin of the effective operators, 
we find that all powers can appear in (\ref{intro1}). Thus, it is possible to have larger corrections to the effective action in the second case compared to the first. 

In the one-goldstino couplings to matter, we find that one consequence of the higher-dimensional operators is to correct the usual simple relations between goldstino interactions and the soft terms. This is similar to what happens, for example, in the context of chiral symmetry breaking, where the pion-nucleon interactions receive corrections analogous to (\ref{intro1}).

The paper is organized as follows. In section \ref{framework} we present a superfield formulation of low scale SUSY breaking within the framework of the MSSM, both in the case where the dynamical sgoldstino scalar is present and in the case where it is not. We also motivate the need to extend the non-linear MSSM by higher dimensional operators. We then discuss two different parametrizations for the couplings of the goldstino multiplet to MSSM fields. In section \ref{EffectiveActionLowScale} we provide the most general deformation of the non-linear MSSM Lagrangian up to dimension six operators. By using appropriate field redefinitions, the redundant operators are eliminated and the Lagrangian is brought to its irreducible form. In section \ref{phenoeffects} we proceed with an analysis of the phenomenological consequences of this irreducible Lagrangian, focusing on Higgs couplings to gauge bosons and fermions, goldstino couplings with monophoton$+\MET$ signatures as well as four-fermion contact interactions. In section \ref{piongoldstino}, we provide a discussion on the analogy between pion-nucleon interactions and goldstino-Higgs-neutralino interactions.


\section{Theoretical framework: The MSSM and SUSY breaking} \label{framework}


The Lagrangian of the Minimal Supersymmetric Standard Model (MSSM) contains operators of the type
\be\label{sp}
S^\dag S\, \Phi^\dag \Phi|_D\,,\quad \mu S \Phi^2|_F\,,\quad S \Phi^3|_F\,,\quad SW^\alpha W_\alpha|_F \ ,
\ee
which deliver the soft breaking terms. In \eqref{sp}, $S=\theta^2 m_{soft}$ is a spurion, $\Phi$ is a matter chiral superfield and $W_\alpha$ is a gauge field strength superfield.  The MSSM Lagrangian is generally considered to be the low energy description of the theory at the SUSY breaking scale $f$, which in perturbative settings typically is in the range $10^{5}\,$GeV $< \sqrt{f} < 10^{12}\,$GeV.
 For low values of $f$, the gravitino, whose mass scales as $f/M_{Planck}$, is effectively massless and can, to a good approximation, be described in terms of its longitudinal component, the goldstino \cite{cddfg}. Furthermore, the superpartner of the goldstino, the sgoldstino \cite{sgold,Espinoza,pr,Bellazzini:2012mh,Petersson:2012nv}, can acquire a mass comparable to the soft scale. The goldstino $G$ and the sgoldstino $x$ are described in terms of a chiral superfield $X=x+\theta \sqrt{2}G+\theta^2 F_X$, whose auxiliary component $F_X$ acquires a vacuum expectation value (VEV) that breaks SUSY.

A simple way to incorporate the couplings of the goldstino and sgoldstino to the MSSM fields is to replace the spurion $S$ in \eqref{sp} by the goldstino multiplet $X$:
\be\label{sD}
{c_1\over M^2}X^\dag X\, \Phi^\dag \Phi|_D\,,\quad {c_2\mu\over M}X \Phi^2|_F\,,\quad {c_3\over M} X \Phi^3|_F\,,\quad {c_4\over M}XW^\alpha W_\alpha|_F \ ,
\ee
where the coefficients $c_i$  depend on the details of the SUSY breaking mechanism. For example, the soft scalar masses can be written as $m_0^2 = -c_1 f^2/M^2 $. After such replacements, the MSSM Lagrangian becomes,
\bea\label{Lgg1}
&&{\cL_{XMSSM}} = \int d^4 \theta \left[ \sum_{i=1,2} (1-\frac{m_i^2}{f^2}  X^{\dagger} X) \ H_i^{\dagger} e^{V_i} H_i +\!\!\!\!\!\!\!\sum_{I=Q,U,D,L,E} \!\!\!(1-\frac{m_I^2}{f^2}  X^{\dagger} X) \ \Phi_I^{\dagger} e^{V_I} \Phi_I  \right] \nonumber\\
&& + \int d^2 \theta \Bigg[ \left(\mu+{B\over f}X\right)\, H_1H_2+\left(1+{A_U\over f}X\right)H_2QU^c+\left(1+{A_D\over f}X\right)QD^cH_1\nonumber
\\
&&+\left(1+{A_E\over f}X\right)LE^cH_1+\sum_{a=1}^3 \frac{1}{16g_ak}\left(1+{2M_a\over f}X\right)Tr[W^\alpha W_\alpha]_a\Bigg] +h.c.  \label{1} \, \ ,
\eea
where the coefficients have been chosen so that the VEV of the auxiliary $F_X$ reproduces the soft masses. The dynamics of the goldstino multiplet can be described in terms of a simple Polonyi model,
\be\label{L_X}
\cL_X=\int d^4\theta \left(1-{m_X^2\over 4f^2}X^\dag X\right)X^\dag X+ \Big\{\int d^2\theta \ f X+h.c.\Big\} \ ,
\ee
where the quartic Kahler operator provides a soft mass for the sgoldstino.

Note that we can restrict ourselves to the case where the sgoldstino is heavy and integrated out, e.g.~when studying processes in the far infrared limit. It has been shown that a universal way to couple a single goldstino to MSSM supermultiplets, without abandoning the convenient language of superfields, is by employing the same goldstino superfield $X$ and imposing the constraint $X^2=0$ \cite{SK,Rocek:1978nb,Lindstrom:1979kq,Casalbuoni:1988xh}. This constraint eliminates the scalar degree of freedom by replacing it by $(GG)/ 2F_X$. Thus, the extension of the MSSM  in terms of either a goldstino only, or a full goldstino supermultiplet, is merely a matter of imposing the constraint $X^2=0$ or not\footnote{The justification and limitations of the constrained superfield formalism is discussed in
\cite{gero}.}.

Apart from the appearance of new light degrees of freedom, low scale SUSY breaking models differ from the MSSM also in other ways. In particular, in an effective field theory (EFT) analysis of models where the SUSY breaking scale is of the order of a few TeV, a set of additional higher-dimensional operators becomes phenomenologically relevant. These new operators are usually neglected in models of perturbative SUSY breaking for the following two reasons:

First, because they generally appear at higher order in $M^{-1}$ with respect to the leading terms (\ref{sD}). Perturbative SUSY breaking models allow for the scale $M$ to be as low as around $50$ TeV, in order to keep $m_{soft}$ at around the TeV scale. Terms with a higher suppression in $M^{-1}$ are therefore negligible with respect to the dominant ones. However, in the case where {\it both SUSY breaking and mediation} to the MSSM arises from strongly coupled physics, the bounds on the SUSY breaking scale $\sqrt{f}$ and the higher-dimensional operator suppression mass scale $M$ are taken directly from experimental constraints and can be as low as a few TeV. In such a scenario, the contributions to processes from the higher-dimensional operators have only a mild suppression compared to the leading ones and they can be particularly significant for couplings that are small in MSSM.

Second, because their goldstino-free components are `hard breaking' terms. However, such terms do not destabilize any mass hierarchy \cite{Espinoza,Martin}. This can be understood by the argument that since the superpotential is not perturbatively renormalized, the only term that could involve a parameter with a positive mass dimension, that could be pushed to the cutoff scale by quantum corrections, is a linear term in the Kahler potential, but such a term is irrelevant in the global SUSY limit.

 A systematic study of the couplings, beyond those in \eqref{sD}, and their consequences, requires a complete classification of all couplings between $X$ and MSSM superfields. The standard EFT procedure is to write down all couplings up to a certain dimension set by the accuracy that we want to achieve. This will produce a large set of operators, many of which will be related to others by field redefinitions or by using equations of motion. Once this redundancy is removed, one can study their phenomenological consequences.

\subsection{Coupling the SUSY breaking sector to the MSSM}
\label{couplings}

As will be defined below, there are two different frameworks one can use in order to  parametrize the couplings of the goldstino
to the MSSM fields in terms of an effective Lagrangian. As a consistence condition, we will require that the effective action
reproduces the standard MSSM with soft breaking terms in the decoupling limit $f \rightarrow \infty$, with fixed values of the soft terms.

i)  In the first case, the Lagrangian coupling the SUSY breaking sector, parametrized by $X$, to MSSM has manifest SUSY, with SUSY
being broken spontaneously at a scale $f$. In this case there is a SUSY messenger sector that mediates interactions between $X$ and the MSSM by integrating out heavy states with a characteristic mass scale $M$.  The theory is in general weakly coupled if $\sqrt{f}, M \geq 50$ TeV and strongly coupled for lower values
of $M$. Since messengers are supersymmetric before coupling to the SUSY breaking sector, all induced operators are manifestly supersymmetric. Standard examples of gravity and gauge mediation are weakly coupled examples in this class.
All higher-dimensional operators in the theory are suppressed by the scale $M$. SUSY is linearly realized and hence the goldstino superfield  contains the elementary  sgoldstino scalar as a propagating degree of freedom \cite{sgold}. A phenomenological analysis within this framework in MSSM was recently performed in \cite{pr,Bellazzini:2012mh,Petersson:2012nv}. At energies  well below the sgoldstino mass, whose value is model-dependent, the sgoldstino can be integrated out, implying that SUSY becomes non-linearly realized in the goldstino multiplet.

ii) In the second case, we make no assumptions about how the SUSY breaking sector couples to the MSSM. The Lagrangian contains the SUSY breaking scale $f$ and the cutoff scale $\Lambda$, where the latter can be related to $f$. In this case, SUSY is always non-linearly realized in the goldstino multiplet $X$ by imposing the constraint $X^2=0$. The sgoldstino is absent as an independent (elementary) degree of freedom and instead replaced by the goldstino bilinear $(GG)/2F_X$. This is the framework discussed in \cite{Antoniadis:2010hs}.

Let us start with case i) above and write the effective Lagrangian in the following form,
\be
{\cal L} \ = \ {\cal L}_X + {\cal L}_{MSSM} + {\cal L}_{soft} + {\cal L}_{hdo} + {\cal L}_{corr} \ , \label{i1}
\ee
where
${\cal L}_X$ and $\ {\cal L}_{MSSM}$ are the SUSY breaking sector Lagrangian and the supersymmetric part of the MSSM action, respectively, whereas,
\bea
&& {\cal L}_{soft} \ = \ - \frac{c_{XQ}}{M^2} \int d^4 \theta \ (X^{\dagger} X) (Q^{\dagger} Q)  \ + \ \cdots \ , \nonumber \\
&& {\cal L}_{hdo} \ = \ -\frac{c_{XX}}{M^2} \int d^4 \theta \ (X^{\dagger} X)^2 - \frac{c_{QQ}}{M^2} \int d^4 \theta \ (Q^{\dagger} Q)^2
+ \cdots \ , \label{i2} \\
&& {\cal L}_{corr} \ = \  -\frac{ c_u}{M^2} \int d^4 \theta \ X^{\dagger} Q U H_1^{\dagger} \ -
\  \frac{d_n}{M^{2 + 2n}} \int d^4 \theta \ (X^{\dagger} X) ({\bar D}^2 {\bar X})^n (Q^{\dagger} Q) +  \cdots ~.\nonumber
\eea
The dots $\cdots$ in ${\cal L}_{soft}$ denote the SUSY operators that give rise to the remaining MSSM soft terms, in ${\cal L}_{hdo}$ they denote
additional higher-dimensional operators which do not involve $X$,  while in ${\cal L}_{corr}$, they denote generic higher-dimensional operators which involve $X$ and provide
corrections to Yukawas, soft terms and other couplings in MSSM. The scalar soft mass terms are given by
\be
m_Q^2 \ = \  c_{XQ} \ \frac{f^2}{M^2} \ . \label{i3}
\ee
In fact, all the soft terms have the structure $m_{soft} \sim \msb = f/M$. The decoupling limit $f \rightarrow \infty$ is therefore defined
with a fixed value of the ratio $f/M$, which can, for example, correspond to the standard rigid limit in supergravity.  From the relations of the kind given in \eqref{i3} we see that the higher-dimensional operators ${\cal L}_{hdo}$ in \eqref{i2}
are suppressed by the appropriate powers of $\msb^2 / f^2$, whereas the corrections to the MSSM couplings, arising from ${\cal L}_{corr}$ in \eqref{i2}, scale as,
\be
\delta y_u = c_u \frac{f}{M^2}\sim \frac{\msb^2}{f} \quad , \quad \delta m_Q^2 = d_n \left(\frac{-f}{M^2}\right)^n \frac{f^2}{M^2} \sim
\left(\frac{\msb^2}{f}\right)^n \ \msb^2 \ . \label{i4}
\ee
If we assume that all dimensionless coefficients $c$ (and $d$), such as those in (\ref{i3}) and \eqref{i4}, are of order one, then the induced SUSY breaking scale $\msb$ in the equations above can be replaced by $m_{soft}$. This is not necessarily the case for perturbative models, in which these dimensionless coefficients correspond to loop factors. However, we are not aware of any perturbative realization, which is phenomenologically viable, for which $\sqrt{f} < 50\,$TeV. Therefore, for such low values of $\sqrt{f}$, we have in mind models that involve some strong dynamics. In such models, the dimensionless coefficients can indeed naturally be of order one (or $4\pi$).

In case ii), it was argued in \cite{SK} that any goldstino coupling should appear in the combination $(m_{soft}/f) X $. The operators providing the soft terms are multiplied by such factors, whereas other higher-dimensional operators are further suppressed by
the appropriate power of the cutoff scale $\Lambda$. The suppression of the goldstino multiplet by $(m_{soft}/f)$ ensures the validity of the effective operator expansion. The analog of the operators in  (\ref{i2}) now reads,
\bea
&&\hspace{-0.7cm} {\cal L}_{soft} \ = \ - \frac{m_Q^2}{f^2} \int d^4 \theta \ (X^{\dagger} X) (Q^{\dagger} Q)  \ + \ \cdots \ , \nonumber \\
&&\hspace{-0.7cm} {\cal L}_{hdo} \ = \  - \frac{c_{QQ}}{\Lambda^2}
\int d^4 \theta \ (Q^{\dagger} Q)^2
+ \cdots \ ,  \label{i5} \\
&&\hspace{-0.7cm} {\cal L}_{corr} \ = \ - \frac{c_u}{\Lambda} \frac{\msb}{f }\!\!\! \int d^4 \theta  X^{\dagger} Q U H_1^{\dagger} - \left(\frac{\msb}{f}\right)^{n+2}\!\frac{d_n}{\Lambda^{n}}\!\! \int d^4 \theta \ (X^{\dagger} X) ({\bar D}^2 {\bar X})^n (Q^{\dagger} Q) +  \cdots
\nonumber
\eea
The precise value of the cutoff scale $\Lambda$ depends on the details of the microscopic theory. For the parametric value $f/\msb \sim \Lambda > \sqrt{f}$, we recover the corrections (\ref{i4}) of case i), with $M$ replaced by $\Lambda$. However, in models involving strongly coupled physics, we do not expect $\Lambda^2$ to be larger than around $4\pi f$ \cite{luty}. Instead, we would expect  the cutoff scale of the theory to be of the same order as the  SUSY breaking scale, i.e.~$\Lambda \sim \sqrt{f}$. In this case the corrections to the MSSM couplings have  the following structure,
\be
\delta y_u = \frac{c_u}{\Lambda} \frac{\msb}{f }f\sim \frac{\msb}{\sqrt{f} } \,,\quad \delta m_Q^2 = d_n\! \left(\frac{\msb}{f}\right)^{n+2}
\frac{(-f)^{2+n}}{\Lambda^n} \sim
\left(\frac{\msb}{\sqrt{f}}\right)^n\!  \msb^2 \ . \label{i6}
\ee
By comparing \eqref{i4} and \eqref{i6} we see that, under the assumption $\Lambda \sim \sqrt{f}$, the corrections to MSSM couplings are larger in case ii), compared to case i).  In both cases it is clear that sizable corrections to couplings are possible only for low scale SUSY breaking, $\sqrt{f} \sim\,$TeV. Moreover, in case i) it is not possible to decouple the scale of the higher-dimensional operators $M$ from $f$, since in this case soft terms are severely suppressed compared to $f$. Therefore, in case i), for self-consistency of effective field theory, it is necessary to have $M$ larger than $\sqrt{f}$ which, in turn, is larger than the induced SUSY breaking scale $\msb$,
\be
\msb \ \lesssim \ \sqrt{f} \ \lesssim \ M \ . \label{i7}
\ee
The arguments above extend straightforwardly to the whole Lagrangian (\ref{i1}). Note that, if the higher dimensional operators which are independent of $X$ are not generated in the model under consideration, then, for example, $c_{QQ}=0$ in \eqref{i5}.

The corrections to the MSSM couplings are proportional to the appropriate
power of $\msb^2/f$ in case i) and  $\msb/\sqrt{f}$ in case ii), when the cutoff $\Lambda \sim \sqrt{f}$. For soft masses in the TeV range and SUSY breaking scale $\sqrt{f} > 10$ TeV, these corrections are negligible. However, we are interested in scenarios in which the SUSY breaking scale is below 10 TeV. In this case, the (moderate) suppression in the  operators ${\cal L}_{hdo} + {\cal L}_{corr}$ can easily be compensated by particular values of MSSM parameters, such as the angles $\alpha$ and $\beta$ appearing in the MSSM Higgs sector or the mixing angle determining the LSP composition. Moreover, some MSSM parameters are ``anomalously" small, such as the quartic Higgs self-coupling, the Yukawa couplings (apart from the top Yukawa) and the Higgs coupling to photons.
Therefore, these couplings are particularly sensitive to the corrections arising from the higher-dimensional operators.
We will discuss several explicit examples of the effects of the subleading operators in the following sections of the paper.

Before analyzing the effects of higher-dimensional operators, let us end this section with a comment on their contribution to the tree level mass of the lightest Higgs boson. They were
discussed within the pure MSSM plus the goldstino multiplet, in the constrained formalism \cite{Antoniadis:2010hs} and with the full
unconstrained goldstino supermultiplet in \cite{pr}. It was noticed in these papers that the dynamics of the goldstino multiplet itself provides a correction to the quartic Higgs self-coupling of the generic type $\delta \lambda \sim m_{soft}^4/f^2$. These corrections have the
virtue that they are model-independent, i.e.~they only depend on MSSM soft terms (and the $\mu$-term, which is related to the soft terms via the electroweak breaking conditions).
We can also add model-dependent higher dimensional operators, with the structure,
\be
{\delta \cal L} \ = \ \ - \frac{c_\lambda}{M^4} \int d^4 \theta \ (X^{\dagger} X) (H_i^{\dagger} H_i)^2 + \cdots  \label{c1}
\ee
in case i), and
\be
{\delta \cal L} \ = \ \ - \frac{c_\lambda}{\Lambda^2}\frac{\msb^2}{f^2 } \int d^4 \theta \ (X^{\dagger} X) (H_i^{\dagger} H_i)^2 + \cdots  \label{c2}
\ee
in case ii). In (\ref{c1}), (\ref{c2}) the dots represent all other operators in the Higgs sector compatible with gauge invariance.
In terms of case i), the corrections from \eqref{c1} are of the order $\delta \lambda \sim m_{soft}^4/f^2$, i.e.~the same order as the ones discussed
in \cite{Antoniadis:2010hs,pr}.
However, in terms of case ii), for $\Lambda \sim \sqrt{f}$, the corrections from \eqref{c2} are  of the order
$\delta \lambda \sim m_{soft}^2/f$, which are larger than the model-independent corrections generated by the goldstino multiplet dynamics. The latter correction becomes similar to the ones from the goldstino dynamics if $\Lambda \sim f/\msb$. Thus, the corrections to the quartic Higgs self-coupling are in general dominated by the model-dependent terms in case ii) of the constrained superfield formalism,  and they can be large enough in order to raise the Higgs mass to the experimentally measured value at ATLAS and CMS \cite{HiggsDis}.

\section{Effective action of low scale SUSY breaking}\label{EffectiveActionLowScale}

In the previous section we have argued that scenarios in which the SUSY breaking scale is not far above the MSSM soft parameters have two main consequences for SUSY model building. First, the particle spectrum is extended by an almost massless gravitino and possibly a sgoldstino scalar. Second, the MSSM interactions in such scenarios receive significant corrections from the higher-dimensional effective operators involving the goldstino supermultiplet and MSSM superfields. The effective Lagrangian consists of the renormalizable operators, supplemented by an expansion in powers of the inverse cut-off scale of the higher dimensional operators. Provided the suppression scale is the highest energy scale of the theory, the higher order operators will be subleading, unless the corresponding first order terms are absent or relatively small. According to the accuracy that we want to achieve, we choose then to disregard all operators above a certain dimension.

If we choose to expand up to dimension-6 operators, the effective Lagrangian becomes,
\be\label{L_TOT2}
\cL=\cL_X+\cL_{coupl}+\cL_6+\cO(M^{-3})
\ee
where $\cL_X$ is given by (\ref{L_X}), $\cL_{coupl}$ is
\bea\label{L_TOT3}
\cL_{coupl}&=&\int d^4\theta \Biggl\{ \, \cZ_i(X,X^\dag)H_i^\dag e^{V_i} H_i+\cZ_I(X,X^\dag)\Phi_I^\dag e^{V_I} \Phi_I+\big(\cA(X,X^\dag)H_1H_2+h.c.\big)\nonumber
\\
&+&{1\over M}\Big[ \cS_u(X,X^\dag)H_1^\dag e^{V_1} QU^c+\cS_d(X,X^\dag)H_2^\dag e^{V_2} QD^c+\cS_e(X,X^\dag)H_2^\dag e^{V_2}LE^c\nonumber
\\
&+&\, \cC_1(X)\nabla H_1 \nabla H_2 +\cC_2(X)H_1\nabla^2 H_2+\cC_3(X,X^\dag) \nabla^2H_1H_2+h.c. \Big] \Biggr\} \nn
\\
&+&\Biggl\{ \int d^2\theta\Bigg[ \cF_h(X)H_1H_2+\cF_u(X,\ov{D}^2X^\dag)H_2QU^c+\cF_d(X,\ov{D}^2X^\dag)QD^cH_1\!\nonumber
\\
&+&\cF_e(X,\ov{D}^2X^\dag)LE^cH_1+\cF_w^a(X,\ov{D}^2X^\dag)W^aW^a\nn
\\
&+&{1\over M}\Big(\cT_1(X) (H_1H_2)^2+\cT_2(X)QD^cQU^c+\cT_3(X)LE^cQU^c \Big)\Bigg]+h.c. \Biggr\}
\eea
and finally, $\cL_6$ consists of dimension-6 operators that do not involve the goldstino multiplet. Its explicit form is given in Appendix \ref{appendixL6}. $M$ is the characteristic scale of the dynamics that generates the effective couplings, e.g. the scale of the messenger fields. In \eqref{L_TOT3}, $i=1,2$, $I=Q,U^c,D^c,L,E^c$ and $a=1,2,3$ while all gauge and flavor indices are suppressed. For compactness, we have denoted $\Phi_Q=Q$, $\Phi_{U^c}=U^c$, and so on, in the kinetic operators of the matter fields. Also, $\nabla_\alpha H_i = e^{-V_i}D_\alpha e^{V_i}H_i$ denotes the gauge covariant superderivative. $\cZ_{i,I}$, $\cA$, $\cS_{U,D,E}$, $\cC_{1,2,3}$, $\cF_{h,k}$ and $\cT_n$ are general functions of $X$, $X^\dag$ and their derivatives. After integration by parts, they are reduced to,
\bea\label{Xfunctions}
\cZ_{i,I}(X,X^\dag)&=&1+\left({\zeta_{i,I} X\over M}+{\zeta_{i,I}'' X^2\over M^2}+ {\zeta_{i,I}'''D^2X\over M^2}+h.c.\right)+{\zeta_{i,I}' X^\dag X\over M^2} \ , \nonumber
\\
\cA(X,X^\dag)&=&{X^\dag\over M}\left(a+{a' X\over M}+{a'' X^\dag\over M}\right)\,,\quad \cS_i(X,X^\dag)=s_{0,i}+{s_iX\over M}+{s'_i X^\dag\over M} \ , \nn
\\
\cC_1(X)&=&{c_1X\over M}\,,\quad \cC_2(X)={c_2X\over M}\,,\quad \cC_3(X,X^\dag)=c_{0,3}+{c_3X\over M}+{c_3'X^\dag\over M} \ , \nn
\\
\cF_h(X)&=&\hat{\mu}+f_hX+{f'_hX^2\over M}+{f''_hX^3\over M^2} \ , \nn
\\
\cF_k(X,\ov{D}^2X^\dag)&=&\hat{y}_k+{f_kX\over M}+{f'_kX^2\over M^2}+{f''_k\ov{D}^2X^\dag\over M^2}\,,\quad k=u,d,e,w \ , \nn
\\
\cT_n(X)&=&t_{0,n}+{t_{n}X\over M}\,,\quad n=1,2,3\, \ .
\eea
Let us make the following comments:

1) The Lagrangian is presented with the effective operators explicitly suppressed by the UV scale $M$, according to case i) of subsection \ref{couplings}. However, as explained in that subsection, interpretation according to case ii) is easily achieved by an appropriate rewriting of the  coefficients.

2) The operators $X^\dag H_1H_2|_D$ and $X^\dag XH_1H_2|_D$ are identical to $ \mu H_1H_2|_F$ and $XH_1H_2|_F$ when $X$ is a spurion, as in the Giudice-Masiero mechanism. In our case $H_1H_2|_F$ is contained in $ F_X^\dag \ov{\theta}^2H_1H_2|_D$ upon replacing $F_X^\dag=-f$. However, the two supersymmetric operators are not identical since the $D$-term operator also includes goldstino couplings, as well as the field $F_X$, rather than its VEV. Therefore, we  keep both these operators in the effective expansion. However, as we will see below, these operators are indeed related to each other via field redefinitions.

3) For the same reason we keep also operators that contain $D^2X$. For example, the operator $\zeta_i'''D^2X H^\dag_ie^{V_i}H_i|_D/M^2$ contributes to the standard kinetic operator $H_i^\dag e^{V_i}H_i|_D$ by $4\zeta_i'''f/M^2$. This is also mentioned in section \ref{framework} for the operator $X^\dag XQ^\dag Q$. Again, such operators are redefined away by a field redefinition $H^\dag_i\to H^\dag_i+\delta\, D^2XH^\dag_i/M^2$, as we will see below.

4) The parameter $\hat{\mu}$, the Yukawa couplings $\hat{y}_I$ and the gauge couplings $\hat{y}_w^a$ are not the physical $\mu$-parameter, Yukawa and gauge couplings, since the corresponding terms receive contributions from other operators and field redefinitions. The physical couplings are obtained when we restrict to the irreducible Lagrangian, which will be given below.

5) Operators that mix flavor families are constrained by experimental results in flavor changing neutral currents (FCNC). We comply with these results by setting the corresponding coefficients proportional to the SM Yukawa couplings, e.g.~$f_u^{ij}\propto \hat{y}_u^{ij}$, $t_{0,2}^{ijkl}\propto\hat{y}_d^{ij}\,\hat{y}_u^{kl}$ etc.

Lagrangian (\ref{L_TOT2}) is the result of an effective expansion in powers of the inverse cut-off scale, up to dimension-six operators. Such expansions typically contain redundant operators, in the sense that they are not independent. Before we study the phenomenological consequences of the theory, we need to remove these redundant operators. Various methods exist that allow one to extract the irreducible part of the effective Lagrangian. One of them involves the use of field redefinitions. Such transformations allow one  to choose a new fields' basis where the redundant operators do not appear. The S-matrix is invariant under a broad set of field redefinitions \cite{Kamefuchi:1961sb,Bergere:1975tr,Arzt}, so we can use the transformed Lagrangian in order to study the phenomenological consequences. In our case, we first shift the superfield content of the theory by the higher derivative terms,
\bea\label{derivredefs}
&&\hspace{-0.8cm}\delta H_1\!=\!{\delta_1\over M}\ov{D}^2(H_2^\dag e^{V_2}\,i\sigma_2)^T\!+\!{\delta_1'\over M^2}X\ov{D}^2(H_2^\dag e^{V_2}\,i\sigma_2)^T\!+\!{\delta_1''\over M^2}\ov{D}^2(X^\dag H_2^\dag e^{V_2}\,i\sigma_2)^T\!+\!{\delta_1'''\over M^2}\ov{D}^2X^\dag H_1\nonumber
\\
&&\hspace{-0.8cm}\delta H_2\!=\!{\delta_2\over M}\ov{D}^2(H_1^\dag e^{V_1}\,i\sigma_2)^T\!+\!{\delta_2'\over M^2}X\ov{D}^2(H_1^\dag e^{V_1}\,i\sigma_2)^T\!+\!{\delta_2''\over M^2}\ov{D}^2(X^\dag H_1^\dag e^{V_1}\,i\sigma_2)^T\!+\!{\delta_2'''\over M^2}\ov{D}^2X^\dag H_2\nonumber
\\
&&\hspace{-0.8cm}\delta \Phi_I\!=\!{\delta_I'''\over M^2}\ov{D}^2X^\dag \Phi_I\,,
\qquad \delta X={\delta_X'''\over M^2}\ov{D}^2(H_1 H_2)^\dag \ ,
\eea
followed by the higher dimensional terms,
\bea\label{dimenredefs}
&&\hspace{-0.7cm}\delta H_1={1\over M}\left(\epsilon_1XH_1+\epsilon_1^uQU^c \right) +{1\over M^2}\left(\epsilon_1'X^2H_1+\epsilon^{u'}_1XQU^c+\epsilon_1'' H_1H_1H_2 \right) \ , \nonumber
\\
&&\hspace{-0.7cm}\delta H_2={1\over M}\!\!\left(\epsilon_2XH_2\!+\!\epsilon_2^dQD^c\!+\!\epsilon_2^eLE^c \right)\!+\!{1\over M^2}\!\!\left(\epsilon_2'X^2H_2\!+\!\epsilon^{d'}_2XQD^c\!+\!\epsilon^{e'}_2XLE^c\!+\!\epsilon_2'' H_2H_1H_2 \right) \nonumber
\\
&&\hspace{-0.7cm}\delta \Phi_I={1\over M}\epsilon_IX\Phi_I +{1\over M^2}\left(\epsilon_I'X^2\Phi_I+\epsilon_I'' \Phi_IH_1H_2 \right)\,,
\\
&&\hspace{-0.7cm}\delta X={\epsilon_X\over M}H_1H_2\!+\!{\epsilon_X^u\over M^2}H_2QU^c\!+\!{\epsilon_X^d\over M^2}QD^cH_1\!+\!{\epsilon_X^e\over M^2}LE^cH_1\!+\!{\epsilon_X^{w\,a}\over M^2}W^aW^a\!+\!{\epsilon_X''\over M^2}XH_1H_2~.\nn
\eea
In accordance with comment 5) above, the redefinition parameters that carry flavor indices are proportional to the corresponding Yukawa couplings, e.g. $\epsilon_1^u\propto \hat{y}_u$ etc. In this fields' basis, the general functions $\cZ_{i,I}$, $\cA$, $\cS_{u,d,e}$, $\cC_{1,2,3}$, $\cF_{h,k}$ and $\cT_n$, given in eqs. (\ref{Xfunctions}), are denoted by $\tilde{\cZ}_{i,I}$, $\tilde{\cA}$, $\tilde{\cS}_{u,d,e}$, $\tilde{\cC}_{1,2,3}$, $\tilde{\cF}_{h,k}$ and $\tilde{\cT}_n$, with coefficients given in eq. (\ref{tildes1}), (\ref{tildes2}), (\ref{tildes3}), (\ref{tildes4}), (\ref{tildes5}), (\ref{tildes6}), (\ref{tildes7}) and (\ref{tildes8}).
The higher derivative transformations (\ref{derivredefs}) and the higher dimensional transformations (\ref{dimenredefs}) also contribute to higher dimensional operators, by acting on the quartic operator in $\cL_X$. From (\ref{derivredefs}) we get,
\be
\cL^{der}_{new}=-\int d^4\theta \Bigl\{ {\delta_X'''\over M^2}{m_X^2\over 2f^2}X^\dag X^2 D^2(H_1H_2)+h.c. \Bigr\}   
\ee
and from the higher dimensional field redefinitions (\ref{dimenredefs}) we obtain,
\bea
\cL^{dim}_{new}&=&-\int d^4\theta \Bigl\{   {m_X^2\over 2f^2}X^\dag XX^\dag\Big({\epsilon_X\over M}H_1H_2+{\epsilon_X''\over M^2}XH_1H_2+{\epsilon_X^u\over M^2}H_2QU^c\nn
\\
&+&{\epsilon_X^d\over M^2}QD^cH_1+{\epsilon_X^e\over M^2}LE^cH_1+{\epsilon_X^{w\,a}\over M^2}W^aW^a\Big) +h.c. \Bigr\} \ .
\eea
We now use the parameter space of the field redefinitions in order to shift away the redundant operators. This will reduce $\tilde{\cL}_{coupl}$ to the irreducible Lagrangian $\tilde{\cL}_{coupl}^{irr}$. We choose,
\bea\label{irred}
\tilde{\cZ}_{i,I}(X,X^\dag)&=&1+{\tilde{\zeta}_{i,I}' \over M^2}X^\dag X \,,\quad \tilde{\cA}(X,X^\dag)=\tilde{a}'' {X^\dag X^\dag\over M^2}\ \ , \nonumber
\\
\tilde{\cC}_i(X,X^\dag)&=&0\,,\quad\tilde{\cS}_i(X,X^\dag)=\tilde{s}_{0,i}+{\tilde{s}'_i\over M}X^\dag\,,\quad i=u,d,e \ , \nn
\\
\tilde{\cF}_h(X)&=&\tilde{\hat{\mu}}+\tilde{f}_hX+{\tilde{f}'_h\over M}X^2+{\tilde{f}_h''\over M^2}X^3 \ , \nn
\\
\tilde{\cF}_k(X)&=&\tilde{\hat{y}}_k+{\tilde{f}_k\over M}X+{\tilde{f}'_k\over M^2}X^2\,,\quad k=u,d,e,w \ , \nn
\\
\tilde{\cT}_1(X)&=&\tilde{t}_{0,1}+{\tilde{t}_{1}\over M}X\,,\quad\tilde{\cT}_2(X)=0\,,\quad\tilde{\cT}_3(X)=0 \ ,
\eea
where we have assumed that all corrections that scale as $\hat{\mu}/M$ are irrelevant (for coefficients  of order one). The reduction is performed by appropriately fixing the redefinition parameters in order to eliminate the redundant operators, while the coefficients of the rest are simply renormalized. The exact relations between redefinition parameters and Lagrangian coefficients are given in eq.~\eqref{elimrelations}. For a particular choice of parameters, $\tilde{\cL}_{coupl}^{irr}$ takes the form,
\bea\label{Lirr''}
&&\hspace{-0.7cm}{\tilde{\cL}^{irr}_{coupl}} = \cL_{XMSSM}+\int d^4\theta {1\over M} \Big[\tilde{\cS}_u(X^\dag)H_1^\dag e^{V_1} QU^c+\tilde{\cS}_d(X^\dag)H_2^\dag e^{V_2} QD^c\nn
\\
&&\hspace{-0.7cm}+\,\tilde{\cS}_e(X^\dag)H_2^\dag e^{V_2}LE^c +{1\over M}\left(a''X^\dag X^\dag H_1H_2\right)\Big]\nonumber
\\
&&\hspace{-0.7cm}+\int d^2\theta \Bigl\{   \,{X^2\over M^2}\left(\tilde{f}_u'H_2QU^c+\tilde{f}_d'QD^cH_1+\tilde{f}_e'LE^cH_1+\sum_{a=1}^3 \frac{\tilde{f}_w^{a'}}{16g_ak}Tr[W^\alpha W_\alpha]_a\right)\nn
\\
&&\hspace{-0.7cm}+{\tilde{f}_h'\over M}X^2H_1H_2+{\tilde{f}_h''\over M^2}X^3H_1H_2+{\tilde{t}_{0,1}\over M}(H_1H_2)^2+{\tilde{t}_1\over M^2}X(H_1H_2)^2 \Bigr\}+h.c. \ ,
\eea
where we have replaced,
\bea
&&\hspace{-1.2cm}\mu=\tilde{\hat{\mu}}\,,\ {B\over f}=\tilde{f}_h\,,\ {1\over 16g_ak}=\tilde{\hat{y}}^a_w\,,\ {2M_a\over 16fg_ak}={\tilde{f}^a_w\over M}\,,\ y_{u,d,e}=\tilde{\hat{y}}_{u,d,e}\,,\ {A_{u,d,e}\over f}={\tilde{f}_{u,d,e}\over M}\,.
\eea
We are now in position to write the final irreducible Lagrangian. It is given by,
\be\label{Lfinal}
\cL_{irr}=\cL_X+\tilde{\cL}_{coupl}^{irr}+\cL^{der}_{new}+\cL^{dim}_{new}+\tilde{\cL}_6 \ ,
\ee
where the individual parts in the RHS are given above and $\tilde{\cL}_6$ denotes the six-dimensional operators of Appendix \ref{appendixL6}, after the field redefinitions have been applied.

The physical consequences of a low SUSY breaking scale are encapsulated, in a model-independent way, by the above Lagrangian. Among all the phenomenology involved, in the following we focus only on certain aspects relevant for LHC physics. In particular, we explore deviations from the SM in the Higgs couplings to gauge bosons and fermions as well as monophoton$ + E_T\hspace{-0.45cm}/\ \ $ signals. We also comment on bounds on the cut-off scale $M$ coming from LHC bounds on four-fermion interactions. In doing that, we choose to ignore the contribution coming from operators in $\tilde{\cL}_6$ and focus on the phenomenology captured by the effective operators that appear in (\ref{Lirr''}).

For the sake of notation simplicity, from now on we drop the tildes from the redefined parameters in (\ref{Lirr''}).

\section{Phenomenological effects}\label{phenoeffects}
In this section we discuss low-energy signatures and constraints arising from standard MSSM couplings to the goldstino $X$ multiplet and new couplings coming from the ``wrong Higgs Yukawas" in (\ref{Lirr''}). We also discuss how LHC results constrain the coefficients of the effective operators, as well as the UV scale $M$ and the SUSY breaking scale $\sqrt{f}$, by considering four-fermion contact interaction terms. In order to set the stage for the discussion concerning corrections to Higgs couplings, we begin by reviewing how a model-independent parametrization of the relevant Higgs couplings can be done.

\subsection{Parametrization of Higgs couplings}

The renormalizable tree level Higgs couplings which are relevant to Higgs searches can be parametrized in the following way \cite{Hcouplings},
\begin{eqnarray}
\label{Ltree}
\mathcal{L}_{\mathrm{ren}}&=&  -c_t \frac{m_t}{v}h \,t \,\bar{t}-c_c \frac{m_c}{v}h \,c \,\bar{c}-c_b \frac{m_b}{v}h \,b \,\bar{b}-c_\tau \frac{m_\tau}{v}h \,\tau \,\bar{\tau} \nn \\
&&+  c_Z \frac{m_Z^2}{v}h \,Z^{\mu} \,Z_\mu +  c_W \frac{2m_W^2}{v}h \,W^{+\mu} \,W^{-}_\mu \ .
  \end{eqnarray}
In the MSSM, these dimensionless $c$-coefficients are, at tree level, given by,
 \begin{eqnarray}
c_t^{\mathrm{MSSM}} = c_c^{\mathrm{MSSM}}=\frac{\cos\alpha}{\sin\beta}~,~
 c_b^{\mathrm{MSSM}}=c_{\tau}^{\mathrm{MSSM}}= -\frac{\sin\alpha}{\cos\beta}~,~c_Z^{\mathrm{MSSM}}=c_W^{\mathrm{MSSM}}=\sin(\beta -\alpha)
\end{eqnarray}
where $\tan\beta$ is the ratio of the VEVs of the two Higgs doublets and $\alpha$ is the mixing angle of the neutral CP-even scalar mass matrix.
The usual SM couplings are obtained in the MSSM decoupling limit, in which $\alpha\to\beta-\pi/2$, implying that $\cos\alpha\to \sin\beta$, $\sin\alpha\to -\cos\beta$ and hence, the $c$-coefficients become $c_i^{\mathrm{MSSM}}\to c_i^{\mathrm{SM}}=1$.

We also consider the following dimension five operators,
\begin{eqnarray}
\label{Lloop}
\mathcal{L}_{\mathrm{dim5}}  = c_\gamma \frac{\alpha_{\mathrm{EM}}}{8\pi  v}   \, h \,F^{\mu\nu} F_{\mu\nu}+c_{Z\gamma}\, \frac{\alpha_{\mathrm{EM}}}{4\pi \sin\theta_w \,v}   \, h \,Z^{\mu\nu} F_{\mu\nu} + c_g\frac{\alpha_{\mathrm{S}}}{12\pi v}  \,h
\Tr G^{\mu\nu} G_{\mu\nu} \ .
 \end{eqnarray}
By using the vertices in \eqref{Ltree}, we can write the dominant 1-loop contributions to these coefficients in the following way \cite{dim5} (see also \cite{Zgamma} for more discussions concerning the $h\to Z\gamma$ channel),
 \begin{eqnarray}
 \label{coefloop}
c_\gamma^{\mathrm{loop}} &=&  c_W\, \mathcal{A}_{1}(\tau_W)+ c_t\, N_c Q_t^2 \mathcal{A}_{1/2}(\tau_t)    \ , \nn
\\
c_{Z\gamma}^{\mathrm{loop}} &=&   c_W \cos\theta_w\,A_1(\tau_W,\lambda_W)+c_t N_c{ Q_t(2T_3^{(t)}-4Q_t\sin^2\theta_w) \over \cos\theta_w} A_{1/2}(\tau_t,\lambda_t) \ , \nn
\\
c_g^{\mathrm{loop}} &=& \frac{3}{4}\left(  c_t\, \mathcal{A}_{1/2}(\tau_t) + c_b\, \mathcal{A}_{1/2}(\tau_b)  \right) \ ,
 \end{eqnarray}
where $\tau_i=4m_i^2/m_h^2$, $\lambda_i=4m_i^2/m_Z^2$, the number of colors is $N_c=3$, while  the electric charge and the weak isospin of the top quark are, respectively, $Q_t=2/3$ and $T_3^{(t)}=1/2$. We have assumed that all loop-contributions from superpartners and additional Higgs scalars are negligible. The 1-loop form factors are given by,
 \begin{eqnarray}
\mathcal{A}_{1/2}(\tau) & = & 2\tau^{2} [\tau^{-1} +(\tau^{-1} -1)f(\tau^{-1})]\,  \ ,  \nonumber
\\
\mathcal{A}_1(\tau) & = & -\tau^{2} [2\tau^{-2} +3\tau^{-1}+3(2\tau^{-1} -1)f(\tau^{-1})]\,  \ , \nn
\\
\mathcal{A}_{1/2}(\tau,\lambda) & = & I_1(\tau,\lambda)-I_2(\tau,\lambda)\ , \nn
\\
\mathcal{A}_1(\tau,\lambda) & = & 4(3-\tan^2\theta_w)I_2(\tau,\lambda)+\big[ (1+2\tau^{-1})\tan^2\theta_w-(5+2\tau^{-1}) \big]I_1(\tau,\lambda) ~ \nn \ ,
\end{eqnarray}
where,
\bea
I_1(\tau,\lambda)&=&{\tau\lambda\over 2(\tau-\lambda)}+{\tau^2\lambda^2\over 2(\tau-\lambda)^2}[f(\tau^{-1})-f(\lambda^{-1})]+{\tau^2\lambda\over (\tau-\lambda)^2}[g(\tau^{-1})-g(\lambda^{-1})] \ , \nn
\\
I_2(\tau,\lambda)&=& -{\tau\lambda\over 2(\tau-\lambda)}[f(\tau^{-1})-f(\lambda^{-1})] \ ,
\eea
 and,
\begin{eqnarray}
f(x)&=&\left\{
\begin{array}{ll}  \displaystyle
\arcsin^2\sqrt{x} & x\leq 1
\\
\displaystyle -\frac{1}{4}\left[ \log\frac{1+\sqrt{1-x^{-1}}}
{1-\sqrt{1-x^{-1}}}-i\pi \right]^2 \hspace{0.5cm} & x>1 \ ,
\end{array} \right.  \\
g(x)&=&\left\{
\begin{array}{ll}  \displaystyle
\sqrt{x^{-1}-1}\arcsin\sqrt{x} & x\leq 1
\\
\displaystyle \frac{\sqrt{1-x^{-1}}}{2}\left[ \log\frac{1+\sqrt{1-x^{-1}}}
{1-\sqrt{1-x^{-1}}}-i\pi \right]^2 \hspace{0.5cm} & x>1 \ .
\end{array} \right.
\end{eqnarray}

In the MSSM decoupling limit, in which all the $c$-coefficients in \eqref{Ltree} are equal to one, the $c^{\mathrm{loop}}$-coefficients in \eqref{coefloop} become equal to the usual 1-loop coefficients of the SM. In this limit, for $m_h=125.5$ GeV, these 1-loop coefficients take the following values,
\begin{equation}
c_\gamma^{\mathrm{loop}}\to c_\gamma^{\mathrm{SM}} \approx -6.51 \ , \
c_{Z\gamma}^{\mathrm{loop}}\to c_{Z\gamma}^{\mathrm{SM}} \approx 5.47 \ , \
c_g^{\mathrm{loop}}\to c_g^{\mathrm{SM}} \approx 0.98+0.07i \ .
\end{equation}

\subsection{Contributions to $h\rightarrow \gamma\gamma$, $h\rightarrow\gamma Z$ and $gg\rightarrow h$}

By using the parametrization of the Higgs couplings outlined in the previous section, let us now consider how sgoldstino-Higgs mixing, arising from $\cL_{XMSSM}$ in \eqref{Lgg1}, affects the dimension-five Higgs couplings to gauge bosons in \eqref{Lloop}; see \cite{diphoton} for discussions on how to enhance the $h\rightarrow \gamma\gamma$ signal from new physics. From (\ref{Lgg1}) we extract the following lowest-order interactions involving the sgoldstino,
\bea
&&{\cal L}\supset x\left(-{m_i^2\over f^2}F_X^\dag \, h_i^\dag F_i+{B\over f} (F_1h_2+h_1F_2)-{M_a\over 4f}(F^{k\, \mu\nu}F^k_{\mu\nu})_a\right)+h.c.\nonumber
\\
&&\quad\quad-|x|^2 \left({m_i^2\over f^2}|F_i|^2+m_X^2\right) \ .
\eea
 In the case where the sgoldstino $x$ is sufficiently heavy we can use its e.o.m., in the zero-momentum limit, in order to integrate it out. By substituting the solution back into the Lagrangian, we obtain,
\be
-{M_a\over 4m_X^2f^2}(F^{k\, \mu\nu}F^k_{\mu\nu})_a\left( m_i^2 h_i^\dag \,F_i+B(F_1h_2+h_1F_2)\right)+h.c.~.
\ee
This allows us to extract the following effective interactions between  the lightest neutral CP-even Higgs $h$ and the gauge  field strengths,
\bea\label{higgs}
&&c_{x}\,\Big[
 (M_{1}\cos^2\theta_w+M_{2} \sin^2\theta_w) h F^{\mu\nu} F_{\mu\nu}+(M_{1}\sin^2\theta_w+M_{2} \cos^2\theta_w) h Z^{\mu\nu} Z_{\mu\nu}\nn
 \\
&&\qquad+2\cos\theta_w\sin\theta_w(M_1-M_2) h Z^{\mu\nu} F_{\mu\nu}+ M_3 \,h\,\Tr \ G^{\mu\nu} G_{\mu\nu}  \Big] \ ,
\eea
where,
\bea
c_{x}&=&{\mu\, v\over 2 f^2 m_X^2}\left( {m_1^2+m_2^2\over 2}\mathrm{cos}(\alpha+\beta)-B\,\mathrm{sin}(\alpha-\beta) \right)\nn
\\
&=&-{\mu\, v\over 2 f^2 m_X^2}\left( \mu^2 \cos(\alpha+\beta)+B\left( \frac{\cos(\alpha+\beta)}{\sin2\beta}+\sin(\alpha-\beta) \right)  \right) \label{cs}
\eea
and with $v_1=v\cos\beta$, $v_2=v\sin\beta$, $v=246\,$GeV. In the last line in (\ref{cs}) we used the MSSM minimization conditions,
\bea\label{EWBC}
m_1^2=-|\mu|^2-B \tan\beta-{g_Y^2+g_2^2\over 8}(v_1^2-v_2^2)+\cO(M^{-1},f^{-1}) \ , \nn
\\
m_2^2=-|\mu|^2-{B\over \tan\beta}+{g_Y^2+g_2^2\over 8}(v_1^2-v_2^2)+\cO(M^{-1},f^{-1}) \, \ .
\eea
 In \eqref{higgs},  we have also used
\begin{eqnarray}
&& h_2\cdot h_1 = h_2^+ h_1^- -h_2^0 h_1^0 \quad , \quad h_i^0 = (1/\sqrt{2}) (v_i+\mathrm{Re}h_1^0+i\mathrm{Im}h_1^0) \ , \nn \\
&& \mathrm{Re}h_1^0 = - \sin\alpha\,h+\cos\alpha\,H \quad , \quad \mathrm{Re}h_2^0 = \cos\alpha\,h+\sin\alpha\,H \ , \nn \\
&& B_{\mu\nu} = \cos\theta_w F_{\mu\nu}-\sin\theta_w Z_{\mu\nu} \quad , \quad W^{(3)}_{\mu\nu} = \sin\theta_w F_{\mu\nu}+\cos\theta_w Z_{\mu\nu} \
\end{eqnarray}
where the gauge bosons of $U(1)_Y$ and (the third component of) $SU(2)_L$, $B_\mu$ and $W^{(3)}_\mu$ have been rewritten in terms of the photon $A_\mu$ and the $Z$-boson $Z_\mu$, with the corresponding field strengths $F_{\mu\nu}$ and $Z_{\mu\nu}$. Note that, since the $h Z^{\mu\nu} Z_{\mu\nu}$-coupling in \eqref{higgs} will be negligible with respect to the $h Z^{\mu} Z_{\mu}$-coupling in \eqref{Ltree} involving the longitudinal components of the $Z$ boson, we will not consider this contribution.

We can now relate the contributions, arising from the operators in \eqref{Lgg1}, to the dimensionless $c$-couplings in \eqref{Lloop} in the following way,
 \begin{equation}
\label{coeff}
c_\gamma = c_\gamma^{\mathrm{loop}} +c_\gamma^{\mathrm{sgold}} \ , \ c_g = c_g^{\mathrm{loop}}+c_g^{\mathrm{sgold}} \ , \
c_{Z\gamma} = c_{Z\gamma}^{\mathrm{loop}}+c_{Z\gamma}^{\mathrm{sgold}} \ ,
 \end{equation}
where,
 \begin{eqnarray}
 \label{coeffB}
c_\gamma^{\mathrm{sgold}} &=& -\frac{4\pi\,v^2\mu}{f^2m_X^2\alpha_{\mathrm{EM}}} (M_{1}\cos^2\theta_w+M_{2} \sin^2\theta_w) \,\Delta \nn \\
c_{Z\gamma}^{\mathrm{sgold}} &=& -\frac{4\pi\,v^2\mu\cos\theta_w\sin^2\theta_w}{f^2m_X^2\alpha_{\mathrm{EM}}}(M_{1}-M_{2})\,\Delta \nn \\
c_{g}^{\mathrm{sgold}} &=&-\frac{6\pi\,v^2\mu}{f^2m_X^2\alpha_{\mathrm{S}}}\, M_{3} \,\Delta ~.
 \end{eqnarray}
 The factor $\Delta$ is given by,
 \begin{equation}
\label{del}
\Delta=\mu^2 \cos(\alpha+\beta)+B\left( \frac{\cos(\alpha+\beta)}{\sin2\beta}+\sin(\alpha-\beta)\right)
\to \mu^2 \sin2\beta
\end{equation}
where, in the last step, we have taken the MSSM decoupling limit, $\alpha\to\beta-\pi/2$. Note that, since we will assume the mixing matrix element between the lightest mass basis scalar $h$ and the sgoldstino gauge basis scalar $x$ to be small, we can, to a good approximation, continue to make use of the MSSM mixing angle $\alpha$ and the  usual MSSM decoupling limit. From the expressions in \eqref{coeffB} it is clear that the sgoldstino mixing contribution to the $h$-couplings vanishes in the limit where we take either the SUSY breaking scale or the sgoldstino soft mass, to be very large.

\begin{figure*}[!t]
\begin{center}
\includegraphics[scale=0.5]{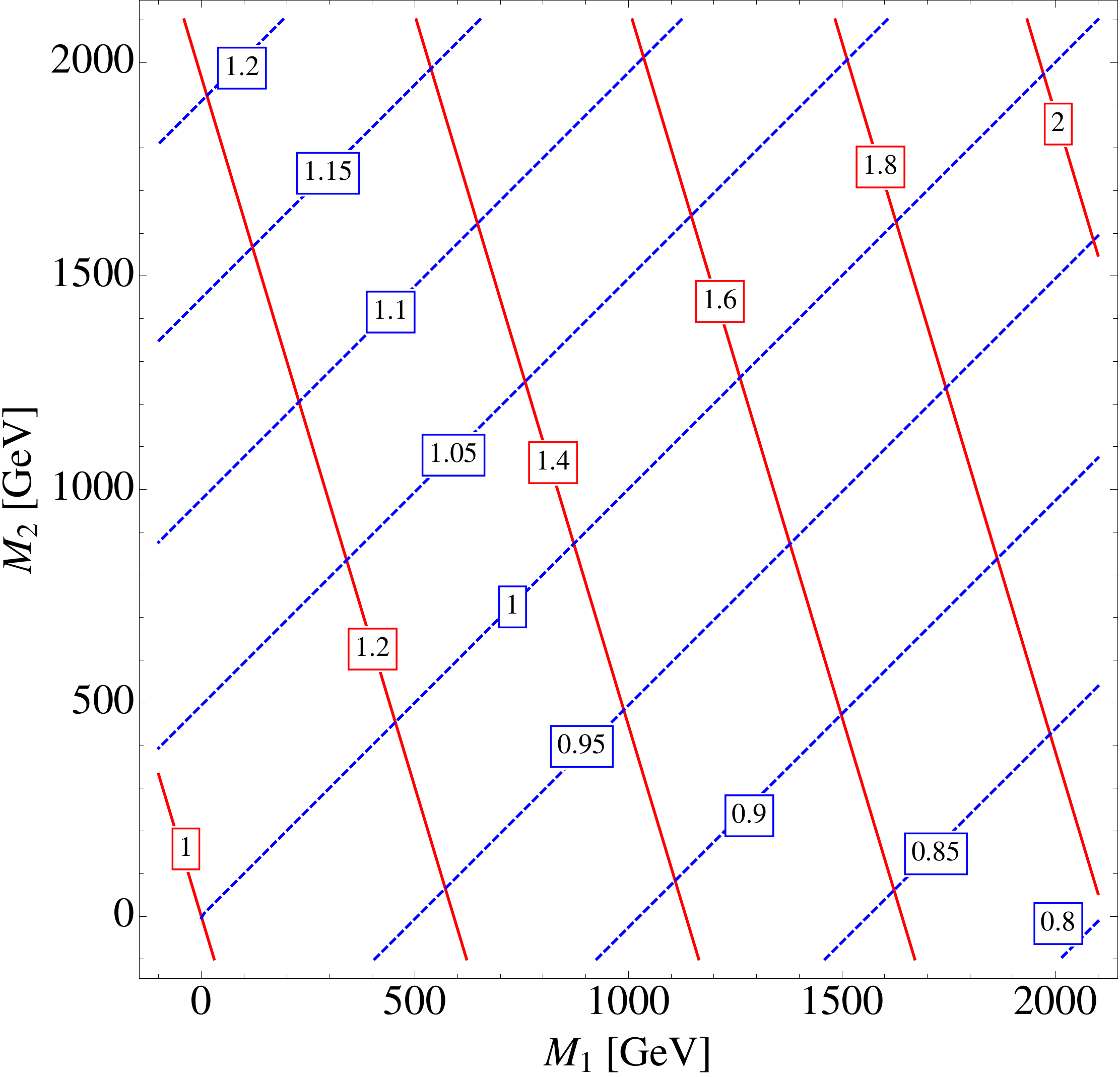}
\end{center}
\caption{
\small The $h\to \gamma\gamma$ and $h\to Z\gamma$  partial decay widths, normalized with respect to their corresponding SM values, i.e.~$\Gamma_{h\gamma\gamma}/\Gamma_{h\gamma\gamma}^{\mathrm{SM}}$ (red solid lines) and $\Gamma_{hZ\gamma}/\Gamma_{hZ\gamma}^{\mathrm{SM}}$ (blue dashed lines), are shown as functions of the bino and wino masses. The values of these normalized partial widths correspond to the maximum values given the assumption that the gluon production cross section does not deviate from the SM value by more than 30\%. The gluino mass $M_3$ has been set to 1 TeV.
}\label{partialwidths}
\end{figure*}

In order for the couplings in \eqref{Ltree} to be close to their corresponding SM value, we consider the MSSM decoupling limit. Since we do not want to modify, for example, the gluon fusion cross section too much, we can use the experimental bound on the gluino mass, which enters the $c_{g}^{\mathrm{sgold}} $ in \eqref{coeffB}, in order to estimate how much the Higgs couplings to $\gamma\gamma$ and $ Z\gamma$ can be enhanced. Say that we do not want gluon fusion cross section to deviate\footnote{Depending on the sign of the gluino mass $M_3$, the sgoldstino mixing contribution to the gluon production cross section can be constructive or destructive.} from SM value by more than around 30\%, i.e. \mbox{$|c_{g}^{\mathrm{sgold}}| \leqslant 0.14\cdot |c_g^{\mathrm{SM}}|$}. This gives us the following relation,
\begin{equation}
\label{ }
\left|-\frac{\mu^3\sin2\beta}{f^2m_X^2}\,  \right|\leqslant 0.14\cdot 0.98\, \frac{\alpha_{\mathrm{S}}}{6\pi\,v^2 \left| M_{3}\right| }
\end{equation}
which we can now insert into $c_{\gamma}^{\mathrm{sgold}} $ in \eqref{coeffB} in order to obtain the following bound,
 \begin{eqnarray}
\label{ }
\left| c_\gamma^{\mathrm{sgold}} \right| 
\leqslant
0.14\cdot 0.98\, \frac{\alpha_{\mathrm{S}}}{6\pi\,v^2 \left| M_{3}\right| }
 \frac{4\pi\,v^2}{\alpha_{\mathrm{EM}}} \left| M_{1}\cos^2\theta_w+M_{2} \sin^2\theta_w \right|
\approx 1.37 \, \left| \frac{M_{12}}{M_{3}} \right|
\end{eqnarray}
where $M_{12}=M_{1}\cos^2\theta_w+M_{2} \sin^2\theta_w$. In terms of the partial decay width, assuming the signs of $\mu$ and $M_{12}$ are such that the sgoldstino mixing contribution is constructive, this implies,
\begin{equation}
\label{gammagamma}
\frac{\Gamma_{h\gamma\gamma}}{\Gamma_{h\gamma\gamma}^{\mathrm{SM}}}=\left| \frac{c_\gamma}{c_\gamma^{\mathrm{SM}}} \right|^2\leqslant \left| \frac{-6.51 - 1.37 \frac{M_{12}}{M_{3}}}{-6.51} \right|^2
\approx  \left| 1 + 0.21 \frac{M_{12}}{M_{3}} \right|^2\,.
\end{equation}
The normalized partial decay width in \eqref{gammagamma} is shown as a function of the bino and wino masses in Figure \ref{partialwidths} (red solid lines), for $M_3$ taken to be 1 TeV. Note that, requiring an amount different than 30 \% in the deviation in the gluon fusion production cross section simply amounts to rescaling appropriately the factor 0.21 in \eqref{gammagamma}.

By the same argument, we can also constrain the $Z\gamma$ channel in the following way,
 \begin{equation}
\label{ }
\left| c_{Z\gamma}^{\mathrm{sgold}} \right|\leqslant 0.14\cdot 0.98\, \frac{\alpha_{\mathrm{S}}}{6\pi\,v^2 \left| M_{3}\right| }
  \frac{4 \pi\,v^2\cos\theta_w \sin^2\theta_w}{ \alpha_{\mathrm{EM}}} \left| M_{2}-M_{1}\right|
\approx 0.28 \left| \frac{M_{2}-M_{1}}{M_{3}} \right| \ ,
\end{equation}
implying that
\begin{equation}
\label{Zgamma}
\frac{\Gamma_{hZ\gamma}}{\Gamma_{hZ\gamma}^{\mathrm{SM}}}=\left| \frac{c_{Z\gamma}}{c_{Z\gamma}^{\mathrm{SM}}} \right|^2\leqslant \left| \frac{5.47 + 0.28 \frac{M_{2}-M_{1}}{M_{3}}}{5.47} \right|^2
\approx  \left| 1 + 0.05 \frac{M_{2}-M_{1}}{M_{3}} \right|^2\, .
\end{equation}
The normalized partial decay width in \eqref{Zgamma} is shown as a function of the bino and wino masses in Figure \ref{partialwidths} (blue dashed lines).
As was already seen in \eqref{coeffB}, the sgoldstino mixing contribution $c_{Z\gamma}^{\mathrm{sgold}}$ vanishes in the case where $M_1=M_2$. Moreover, Figure \ref{partialwidths}  highlights the fact that, in general, we expect a smaller deviation from the SM value in the $h\to Z\gamma$ channel than in the $h\to \gamma\gamma$ channel.

\subsection{Higgs couplings to fermions}\label{Higgsfermions}

Another important aspect of low-scale SUSY breaking are the deviations of the Higgs couplings to SM fermions. Let us consider the following  operators from \eqref{Lirr''},
\bea
\label{wrong}
\cL\supset\!\int d^4\theta\, \Bigg[ {s_u^{\prime} \over M^2} X^\dag H_1^\dag e^{V_1} QU^c+{s_d^{\prime } \over M^2} X^\dag H_2^\dag e^{V_2} QD^c+{s_e^{\prime } \over M^2} X^\dag H_2^\dag e^{V_2}LE^c +h.c. \Big]
\eea
where the flavor indices are suppressed (and as we mentioned earlier, the tildes have been dropped). These ``wrong'' Yukawa operators give contributions to both the fermion masses and the Yukawa couplings for the lightest neutral CP-even Higgs boson $h$ which, together with the usual MSSM contributions, become,
\bea
\label{wrongyuk}
m_u=\left( y_u \sin\beta+{s_u^{\prime } f\over M^2}\cos\beta \right){v\over \sqrt{2}}&,&g_{huu}=\frac{1}{\sqrt{2}}\left( y_u \cos\alpha-{s_u^{\prime } f\over M^2}\sin\alpha \right)\nn \\
m_d=\left( y_d \cos\beta+{s_d^{\prime } f\over M^2}\sin\beta \right){v\over \sqrt{2}}&,&g_{hdd}=\frac{1}{\sqrt{2}}\left( - y_d \sin\alpha+{s_d^{\prime } f\over M^2}\cos\alpha \right)\nn \\
m_e=\left( y_e \cos\beta+{s_e^{\prime } f\over M^2}\sin\beta \right){v\over \sqrt{2}}&,&g_{hee}=\frac{1}{\sqrt{2}}\left( - y_e \sin\alpha+{s_e^{\prime } f\over M^2}\cos\alpha \right).
\eea
In case i) of the discussion in section \ref{couplings}, these couplings are suppressed by a factor $f/M^2 \sim \msb^2/f$ compared to the MSSM couplings. In the case ii), without a dynamical sgoldstino and with a cutoff $\Lambda \sim \sqrt{f}$, these couplings are suppressed as $\msb/ \Lambda \sim \msb/\sqrt{f}$. These extra contributions can have a relevant effect on those Yukawa couplings in the SM (and in the MSSM) that are small.

As we already mentioned in section \ref{EffectiveActionLowScale}, if the ``wrong'' Yukawa operators in \eqref{wrong} introduce flavor structure, in terms of the dimensionless coefficients $s_u^{\prime}$,  $s_d^{\prime}$ and $s_e^{\prime}$, which is not aligned with the SM Yukawa structure, then the operators in \eqref{wrong} can introduce significant contributions to flavor changing processes. Therefore, constraints on these coefficients imply constraints on the dynamics of the sector that generates these operators. Wrong-Higgs Yukawa couplings that are aligned with the ordinary Yukawa couplings deliver $\tan\beta$-enhanced effects in the relation between masses and corresponding Yukawa couplings. Wrong-Higgs Yukawa couplings induced by one-loop radiative corrections have been studied in \cite{wronghiggs}.

It is only away from the decoupling limit that the couplings in \eqref{wrongyuk} can significantly depart from the usual MSSM Yukawa couplings, which are recovered upon setting $s_u^{\prime}=s_d^{\prime}=s_e^{\prime}=0$. In the decoupling limit, where $\cos\alpha\to \sin\beta$ and $\sin\alpha\to -\cos\beta$, the couplings in \eqref{wrongyuk} reproduce the usual SM Yukawas, $g_{huu}\to m_u/v$, $g_{hdd}\to m_d/v$ and $g_{hee}\to m_e/v$, without any dependence on the $s^{\prime}$-coefficients. However, the dependence on the $s^{\prime}$-coefficients does not vanish in the decoupling limit for the Yukawa couplings of the heavy neutral Higgs scalar $H$, given by,
 \bea
\label{heavyhiggs}
g_{Huu}&=&\frac{1}{\sqrt{2}}\left( y_u \sin\alpha+{s_u^{\prime } f\over M^2}\cos\alpha \right)\nn \\
g_{Hdd}&=&\frac{1}{\sqrt{2}}\left( y_d \cos\alpha+{s_d^{\prime } f\over M^2}\sin\alpha \right)\nn \\
g_{Hee}&=&\frac{1}{\sqrt{2}}\left( y_e \cos\alpha+{s_e^{\prime } f\over M^2}\sin\alpha \right)~.
\eea
Hence, in the case where the mass of $H$ is not too large, unless the $s^{\prime}$-coefficients are aligned with the SM Yukawas, flavor changing contributions can arise through the couplings of $H$.

Let us also comment on the fact the SUSY operators in \eqref{Lgg1} which give rise to the soft $A$-terms, also give rise to Yukawa-like interactions of the sgoldstino scalar and the SM fermions. Due to these interactions, through mixing between the sgoldstino and the Higgs scalar, a dependence on the $A$-term parameters is induced for the Yukawa couplings of the lightest (mass-basis) Higgs scalar $h$ \cite{Petersson:2012nv}. In analogy with the dependence on the $s^{\prime}$-coefficients in \eqref{wrongyuk}, also this dependence on the $A$-terms allows one to disentangle the usual MSSM Yukawa relations\footnote{In \cite{Petersson:2012nv}, also the $\tilde{f}_h'$-operator in \eqref{Lirr''} was considered, giving rise to a trilinear scalar interaction that provides mixing contributions. In the present paper we consider the case where $\tilde{f}_h'$ is small and the sgoldstino mixing contributions to \eqref{wrongyuk} are negligible.}. In terms of the Yukawa couplings relevant for Higgs production and decay, while having the possibility to generate significant contributions to the tau and bottom Yukawa couplings, it is more difficult to affect the top Yukawa since its SM value is much greater.

For a study of Yukawa couplings in the context of one Higgs doublet MSSM-like model, see \cite{kehagias}.
\subsection{Monophoton$+\MET$ signatures}

In this section we discuss the final state consisting of a single photon and two goldstinos, giving rise to a monophoton$+\MET$ signature. Depending on the energy range for the transverse momenta of the photon  $p_T^\gamma$, different processes provide the dominant contribution to this final state, which we discuss separately below.

\subsubsection{Processes with a high momentum photon}

\begin{figure}[!t]
\begin{center}
\includegraphics[scale=0.5]{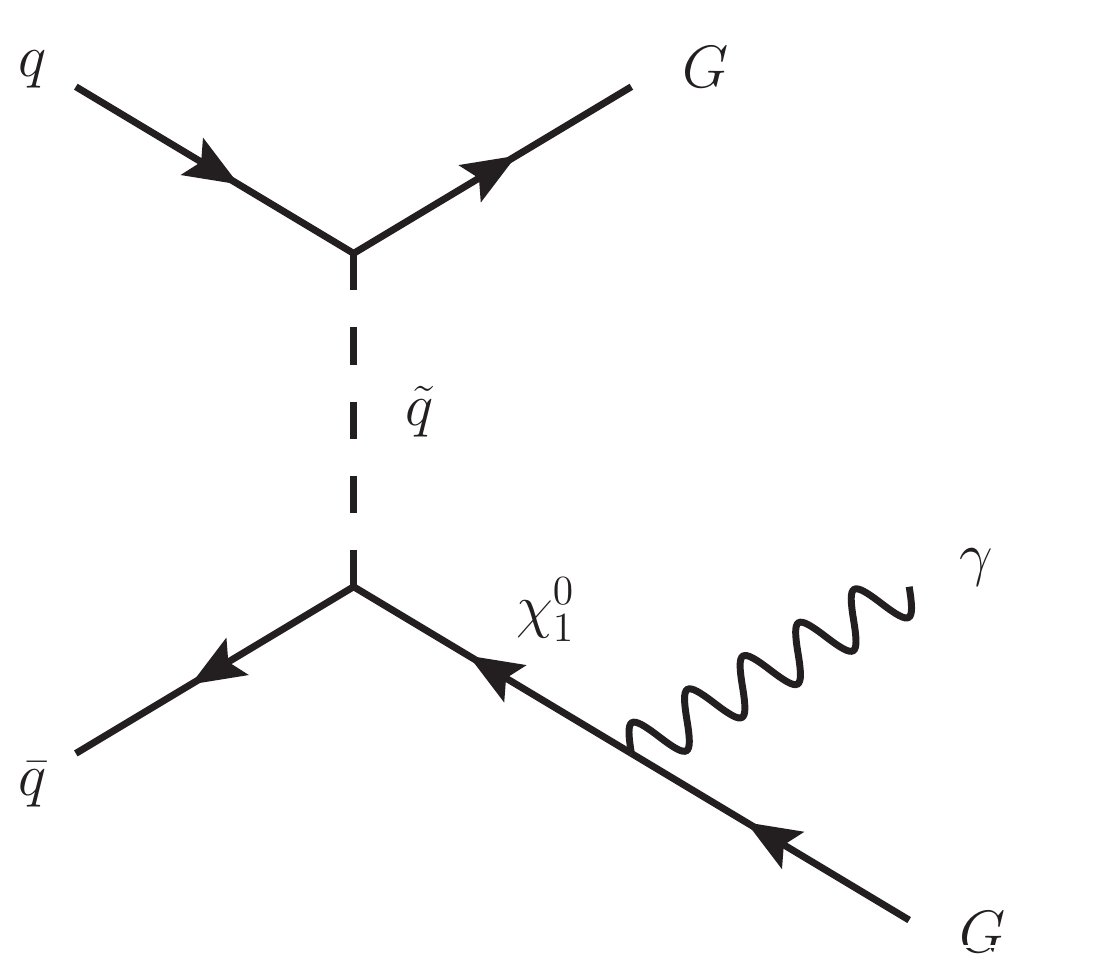}
\end{center}
\caption{
\small
Feynman diagram for the process $q\bar{q}\to \chi_{1} ^{0}G\to \gamma GG$.
}\label{fig:squark}
\end{figure}

In the regime where the $p_T^\gamma$ is high, of the order of around 100 GeV or higher, one contribution to the $\gamma GG$ final state comes from the non-resonant process in Figure \ref{fig:squark}. This process arises from when a quark-squark-goldstino vertex and a quark-squark-neutralino vertex from \eqref{Lgg1} are connected via a t-channel squark exchange \cite{monophoton}. For a low SUSY breaking scale, around the TeV scale, the  neutralino will always decay promptly on collider time scales. In the case where the lightest neutralino mass is  above the mass  of the $Z$or the Higgs boson, it can also decay into a $Z$, or a Higgs, and a goldstino.

The contribution from the process $q\bar{q}\to \chi_{1} ^{0}G\to \gamma GG$, depicted in Figure \ref{fig:squark}, is maximized when the neutralino is gaugino-like since, in this case, the quark-squark-neutralino coupling is given in terms of the gauge couplings. In contrast, in the case where the neutralino is higgsino-like, the quark-squark-neutralino coupling will be given mainly in terms of the  Yukawa couplings.  If, in addition, the t-channel squark is sufficiently heavy, it can be integrated out and the following effective dimension six operators are generated,
\bea
\label{comp2}
&&\cL\supset {y_u\over f}\, \ov{G\psi}{}_2^0\,u u^c+ {y_d\over f}\,\ov{G\psi}{}_1^0\,dd^c+{y_e\over f}\,\ov{G\psi}{}_1^0\,ee^c+h.c.
\eea
These four-fermion vertices give rise to contact interactions for which there is a $q\bar{q}$ pair in the initial state\footnote{For an electron-positron collider, the relevant operator is the third in \eqref{comp2} and an $e^+e^-$ pair appears in the initial state.}, while in the final state there is a goldstino and a neutralino, which promptly decays into another goldstino and a photon (or a $Z$/$h$ boson if kinematically allowed). Note that the sfermion propagator cancels out the dependence on the sfermion mass of the quark-squark-goldstino coupling. Since, to good approximation, only light quarks will appear in the initial state, such as the up and the down quarks, the interactions in \eqref{comp2} will be strongly Yukawa suppressed. Therefore, in the case where the neutralino is higgsino-like, we expect the contribution to the monophoton$+\MET$ final state from the operators in \eqref{comp2} to be small.

However, there are other higher-dimensional operators that contribute to the mono-photon$+\MET$ final state, and their contributions turn out to be maximized for a higgsino-like neutralino. Examples of such operators are the ``wrong'' Yukawa operators in \eqref{wrong}, which give rise to the following four-fermi interactions,
\bea
\label{comp}
&&\cL\supset {s'_u\over M^2}\,\ov{G\psi}{}_1^0\,u u^c+ {s'_d\over M^2}\,\ov{G\psi}{}_2^0\,dd^c+{s'_e\over M^2}\,\ov{G\psi}{}_2^0\,ee^c+h.c.~.
\eea
 Since it is  the higgsinos that are involved in \eqref{comp}, in the case where the neutralino mass is below the $Z$ boson mass, the contributions to the monophoton$+\MET$ final state from the interactions in \eqref{comp} will be maximized when the neutralino is higgsino-like, i.e.~when the neutralino mixing suppression is minimized. In this case, since the branching ratio for the neutralino decay into a goldstino and a photon is equal to one, there is no mixing suppression arising from the neutralino decay part of this process. As was discussed in section \ref{Higgsfermions}, we assume the flavor structure of the $s'$-coefficients in \eqref{comp} to be aligned with the SM Yukawa couplings, however with an undetermined proportionality coefficient.

In agreement with our general discussion in Section \ref{couplings}, the operators in (\ref{comp}) are suppressed compared to the ones in
(\ref{comp2}) by a factor $\msb^2/f^2$. However, as we have discussed, this suppression can be overcome in low-energy SUSY breaking scenarios for which the neutralino is mostly higgsino.

In \cite{Brignole:1998me}, the authors considered the case where both the squarks and the neutralino of the process in Figure \ref{fig:squark} were very heavy and integrated out, such that an effective dimension eight operator was generated (see \cite{Brignole:1997sk} for the analogous analysis relevant for LEP). In their case the neutralino was a pure photino and therefore the prefactor of the dimension eight operator was simply given by the electric coupling over $f^2$, since the photino soft mass of the photino-goldstino-photon vertex was again cancelled by the photino propagator. The lack of an experimental signal in mono-photon$+\MET$ searches at Tevatron \cite{Acosta:2002eq} was used in order to set a constraint on this prefactor, which in their case implied setting a lower bound on $\sqrt{f}$ at around 300 GeV. For a more recent mono-photon$+\MET$ search, performed at the LHC in the context of extra dimensions, see \cite{CMS}.  

In contrast to the case where the neutralino is heavy and integrated out, in which the resulting dimension eight operators contribute to the $2\to 3$ scattering process $q\bar{q}\to \gamma GG$, when the neutralino is light, the dimension six operators in \eqref{comp2} and \eqref{comp} contribute to the  $2\to 2$ process $q\bar{q}\to \chi_1^0 G\to\gamma GG$. As was mentioned above, when the neutralino mass is taken to be above the $Z$ and $h$ boson masses, in addition to the $\chi_1^0\to \gamma G$ decay mode, also  $\chi_1^0\to Z G$ and $\chi_1^0\to h G$ become kinematically accessible. In the limit where the neutralino is sufficiently heavy, in analogy with \cite{Brignole:1998me}, it can be integrated out by using the component interactions of \eqref{Lgg1}, and higher-dimensional effective operators are generated, for which the coefficients depend on the neutralino mixing matrix elements. Hence, the bound on $\sqrt{f}$ in \cite{Brignole:1998me}, for a general neutralino mass and composition,  becomes a bound that also involves the parameters of the neutralino mass matrix.

\subsubsection{Processes with a low momentum photon}\label{Higgsdecay}

\begin{figure}[!t]
\begin{center}
\includegraphics[scale=0.5]{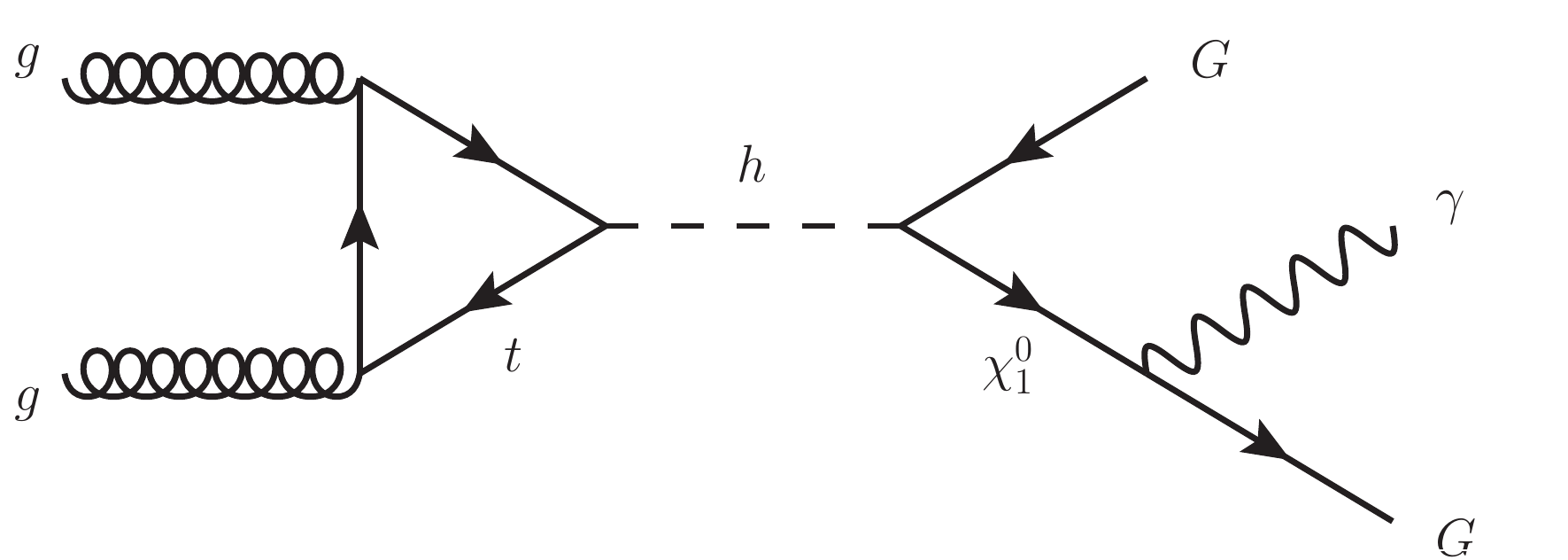}
\end{center}
\caption{
\small
Feynman diagram for the process $gg\to h\to \chi_{1}^{0}G\to\gamma GG$.
}\label{fig:higgs}
\end{figure}

We have so far discussed processes contributing to the monophoton$+\MET$ final state in the phase space regime where the transverse momenta of the photon $p_T^\gamma$ is high, around 100 GeV or higher. However, low scale SUSY breaking scenarios also allow for processes that contribute to the monophoton$+\MET$ final state in the phase space regime where $p_T^\gamma$ is low. One such process is when the lightest Higgs boson $h$ decays into goldstino and a neutralino, which subsequently decays into another goldstino and a photon. Since, at the LHC, the dominant Higgs production mode is the gluon-gluon fusion, the process we are interested in is $gg\to h\to \chi_{1}^{0}G\to\gamma GG$, depicted in Figure~\ref{fig:higgs}.

Since the Higgs is produced resonantly, the process in Figure~\ref{fig:higgs} will only be relevant in the phase space regime where the transverse momenta of the photon $p_T^\gamma<m_h/2$ since the  $p_T^\gamma$ distribution will have a kinematic endpoint at $m_h/2$.
In contrast to the high $p_T^\gamma$ processes discussed above, for which the dominant background is due to $pp\to \gamma Z\to\gamma \nu\nu$, the dominant background in the regime $p_T^\gamma<m_h/2$ is due to $pp\to \gamma j$, where the jet is not reconstructed. A phenomenological analysis of these processes and the relevant backgrounds was done in \cite{Petersson:2012dp} (see also \cite{Djouadi:1997gw,Antoniadis:2010hs,Gherghetta:2011na} for earlier discussions concening the decay $h\to \chi^0_1G$), with the conclusion that, even though the monophoton$+\MET$ final state at low $p_T^\gamma$ is subject to large systematic uncertainties, involving reconstruction efficiencies and energy resolution in the forward region, a counting experiment of signal and background events in a certain phase space region, chosen in order to optimize the signal significance, suggests the possibility of discovery/exclusion already with the data from the 2012 LHC run.

\subsection{LHC bounds on four-fermion contact interactions}

Lagrangian (\ref{Lfinal}) contains four-fermion contact interactions (CI), either in $\tilde{\cL}_6$ given in appendix \ref{appendixL6} or in lower dimensional superfield operators, by integrating out the auxiliary fields or the sgoldstino. The Exotics groups of the ATLAS and CMS collaborations have recently published bounds on certain four-fermion contact interaction (CI) terms \cite{CIATLAS,CICMS} that can be translated into bounds on the coefficients of the corresponding dimension-six operators.

The general form of the CI terms under study is $\eta{c_{CI}\over \Lambda^2}(\ov{\Psi}_{1} \gamma^\mu\, \Psi_{2})(\ov{\Psi}_{3} \gamma_\mu\, \Psi_{4})$ where $\eta=\pm1$ denotes constructive or destructive interference with the SM contribution. The bound on $\Psi_{1,2,3,4}=Q_L=(U_L\ \ D_L)^T$, $c_{CI}=2\pi$ and destructive interference is $\Lambda_{4q}>7.8\,$TeV \cite{CIATLAS}. A similar bound is obtained for CI between 2 quarks and 2 leptons. The combined bound from the dimuon and dielectron channels is $\Lambda_{ql}>10.2\,$TeV (9.4 TeV) for constructive (destructive) interference \cite{CIATLAS}. In both cases, the bounds delivered refer to contact interactions between left-handed quark and lepton fields. Furthermore, bounds on same-sign top pair production ($uutt$) are obtained by same-sign dilepton + jets + MET. The considered effective terms include left-right mixing (notice the 1/2 difference with respect to the general expression above)
\bea
&&{c_{LL}\over 2\Lambda^2}\ov{U}_{L} \gamma^\mu\, T_{L}\ov{U}_{L} \gamma_\mu\, T_{L}+{c_{RR}\over 2\Lambda^2}\ov{U}_{R} \gamma^\mu\, T_{R}\ov{U}_{R} \gamma_\mu\, T_{R}\nn
\\
&-&{c_{LR}\over 2\Lambda^2}\ov{U}_{L} \gamma^\mu\, T_{L}\ov{U}_{R} \gamma_\mu\, T_{R}-{c_{LR}'\over 2\Lambda^2}\ov{U}_{L\,a} \gamma^\mu\, T_{L\,b}\ov{U}_{R\,b} \gamma_\mu\, T_{R\,a}+h.c.
\eea
and the delivered bounds are
\be\label{uuttbounds}
{c_{LL}\over \Lambda_{ut}^2}={c_{RR}\over \Lambda_{ut}^2}<0.35 \,\mathrm{TeV}^{-2}\,,\quad {c_{LR}\over \Lambda_{ut}^2}={c_{LR}'\over \Lambda_{ut}^2}<0.98 \,\mathrm{TeV}^{-2}\,.
\ee

The above results directly translate into bounds on the ratio of the coefficients of the  dimension-six operators (in particular $d_6^{IJ}$, $d_6$, $d_6'$, $d_6''$ and $d_6'''$) over the suppression scale as well as on the quantities that parametrize CI terms that are obtained by integrating out the sgoldstino or auxiliary fields, as we mentioned above. In the following we focus on the contribution from the latter, in particular in the following part of Lagrangian (\ref{Lfinal}):
\bea\label{Luutt}
&&\hspace{-0.7cm}\cL_{irr}\supset \int d^4\theta {1\over M} \Big(s_{0,u}H_1^\dag e^{V_1} QU^c+s_{0,d}H_2^\dag e^{V_2} QD^c+s_{0,e}H_2^\dag e^{V_2}LE^c+h.c.\Big)\nn
\\
&&\hspace{-0.7cm}+ \Bigg(\int d^2\theta\ {A_u\over f}XH_2QU^c+{A_d\over f}XQD^cH_1+{A_e\over f}XLE^cH_1+h.c.\Bigg)\,.
\eea
The operators of the first line above deliver four-fermion CI terms after integrating out the auxiliary fields of the Higgs multiplets. For a sufficiently heavy sgoldstino, the operators in the second line above can also deliver CI terms, after integrating out the sgoldstino in a tree level exchange between two pairs of fermions. 

Since these operators involve interactions between left-handed and right-handed fields, the relevant bounds are the ones on $c_{LR}$ for $uutt$ operators, given in (\ref{uuttbounds}). We have $s_{0,u}^{1\,3}s_{0,u}^{1\,3}/2M^2 < 0.49\,$TeV$^{-2}$ or $M/s_{0,u}^{1\,3}> 1\,$TeV. The same bound on the CI term obtained from the first operator of the second line in (\ref{Luutt}) delivers $v^2\sin^2\beta A_u^{1\,3}A_u^{1\,3}/2m_x^2f^2<0.49\,$TeV$^{-2}$, which for $v=246\,$GeV, $m_x=1\,$TeV and $\sin\beta\simeq1$, delivers $f/A_u^{1\,3}>0.246\,$TeV.

\section{Pion-nucleon interactions vs.~goldstino-Higgs interactions}\label{piongoldstino}

In this section we discuss some general features of spontaneous symmetry breaking and the relation between soft terms and the irreducible Goldstone interactions. We are interested in corrections to goldstino interactions, arising from higher-dimensional operators, and we give an example in terms of the goldstino coupling to a Higgs-higgsino pair, relevant for the discussion in section \ref{Higgsdecay}. In order to highlight the general features, we begin by reviewing these issues in the context of chiral symmetry breaking and pion interactions. 

As a general consequence of spontaneous symmetry breaking, the coupling of the pion-nucleon interaction is proportional to the symmetry breaking parameter, i.e.~the nucleon mass $m_N$, and inversely proportional to the scale of the spontaneous symmetry breaking, which in this case is given by $f_\pi$. Upon considering the most general effective Lagrangian that is invariant under the chiral symmetry, at lowest order in the expansion in terms of pion momentum, the pion-nucleon interaction coupling is given by $g_A m_N/f_\pi$, where $g_A$ is the zero-momentum limit of the matrix element of the axial current between nucleons. This is the famous Goldberger-Treiman relation \cite{Goldberger:1958tr}.   

Note that, if one instead would have truncated the general effective Lagrangian by, for example, neglecting non-renormalizable operators, one would get that $g_A=1$. However, this result would simply be an artifact of the truncation. Since the pion-nucleon interaction is invariant under the chiral symmetry by itself, there is no symmetry that determines the value of $g_A$. Hence, the soft nucleon mass term and the pion-nucleon interaction are two independent operators and $g_A$ can only be fixed by experiments, which measure it to be $g_A^{\mathrm{exp}}\approx 1.26$, see \cite{Weinberg:1996kr} for a detailed discussion concerning these issues and for references to the original work. In addition, at finite pion momentum, there are other, independent, operators which provide correction to the (zero-momentum) Goldberger-Treiman relation.    

In the case of spontaneously broken global SUSY, the associated Goldstone particle is the goldstino fermion and the soft terms for the superpartners provide the symmetry breaking terms. 
In analogy with the pion-nucleon discussion, by considering the most general effective supersymmetric Lagrangian,  at lowest order in the momentum expansion, the goldstino-boson-fermion interaction is given by $g_G m_{soft}^2/f$, where $g_G$ is the zero-momentum limit of the matrix element of the supercurrent between the fermion and the boson. This result is the SUSY version of the Goldberger-Treiman relation \cite{Clark:1996aw}. If we only consider the usual derivative coupling of the goldstino to the supercurrent, we would get that $g_G=1$. Again this result would only be an artifact of our approximation and by considering a more general set of effective operators we would conclude that $g_G$ is a free parameter and not determined by any symmetry. Moreover, operators of higher order in derivatives will correct the (zero-momentum) SUSY Goldberger-Treiman relation. We will now turn to an explicit SUSY example that illustrates the key points of this discussion.  

In terms of the Lagrangian (\ref{L_TOT3}), let us consider the goldstino-Higgs-neutralino coupling. The relevant part of the SUSY Lagrangian for the particular coupling is,
\bea
\label{bla}
\cL
&=&\int d^4\theta \left(1 -{m_1^2\over f^2}X^\dag X\right)H_1^\dag e^{V_1}H_1+\int d^4\theta  \left(1 -{m_2^2\over f^2}X^\dag X\right)H_2^\dag e^{V_2}H_2\nonumber
\\
&+&\left( {B\over f}\int d^2\theta XH_1H_2+{M_i\over 8 g_i^2 k f}\int d^2\theta\, X\, \mathrm{Tr}[W^\alpha W_\alpha]_i +h.c.\right)\nonumber
\\
&+&\Big({c_1\over M^2} \int d^4\theta\,X\, \nabla^\alpha H_1 \nabla_\alpha H_2 +h.c.\Big)\,.
\eea
From the full supersymmetric Lagrangian, we isolate the terms that contribute to the coupling of interest
\bea
\label{Ldecaycomponents}
\cL_{h\rightarrow \chi\,G}
&=&-{1\over f} \big(m_1^2 G\psi_1h_1^{0*}+m_2^2 G\psi_2 h_2^{0*}\big)-{B\over f}\big( G\psi_2h_1^0+G\psi_1h_2^0 \big)\nonumber\\
&&-{1\over f}\sum_{i=1,2}{M_i\over \sqrt{2}}\tilde{D}^a G \lambda^a_i -{1\over \sqrt{2}} \big( g_2\lambda_2^3-g_1\lambda_1 \big)\big( h_1^{0*}\psi_1^0- h_2^{0*}\psi_2^0 \big)\nonumber\\
&&-{2c_1\over M^2} G \big( \partial^\mu h_1\partial_\mu \psi_2+\partial^\mu \psi_1\partial_\mu h_2 \big)\!+\!h.c.
\eea
In \eqref{bla}, the first three operators are those that arise from a supersymmetrization of the supersymmetry breaking soft terms $m_1^2$, $m_2^2$ and $B_\mu$. The leading order goldstino interactions extracted from these three supersymmetric operators, displayed in the first line of \eqref{Ldecaycomponents}, are precisely those one obtains by coupling the goldstino derivatively to the supercurrent. Again, if we only were to consider these operators we would get that the coefficients of these interactions are given by the ratio of the corresponding soft parameter over $f$, i.e.~with proportionality coefficients being equal to one. We will now see that the terms in the second and third lines of \eqref{Ldecaycomponents} provide corrections to the goldstino interactions in the first line, showing that the goldstino interactions and the soft terms are independent operators and there is no symmetry that determines the proportionality coefficients.  

In (\ref{Ldecaycomponents}), the second term of the second line arises from the kinetic terms of the Higgs superfields. We have kept this term since it contributes to the Higgs decay from the goldstino component of $\psi_i$ and $\lambda_i$. The e.o.m.~of $\psi_i$ and $\lambda_i$ from the zero momentum terms give,
\bea
\mu \psi_1^0&=& {1\over f\sqrt{2}}\big( -m_2^2 v_2-Bv_1-{v_2\over 2}<g_2D_2^3-g_1D_1> \big) G+...\ , \nonumber
 \\
\mu \psi_2^0&=& {1\over f\sqrt{2}}\big( -m_1^2 v_1-Bv_2+{v_1\over 2}<g_2D_2^3-g_1D_1> \big) G+...\ , \nonumber
\\
\lambda_1&=&-{1\over \sqrt{2}f}<D_1> G+...\quad , \quad \lambda_2^3 = - {1\over \sqrt{2}f}<D_2^3> G+...
\eea
where
\be
<D_1>={g_1\over 4}(v_1^2-v_2^2),\quad<D_2>=-{g_2\over 4}(v_1^2-v_2^2),\quad <g_2D_2^3-g_1D_1>=-{g^2\over 4}(v_1^2-v_2^2) \ .
\ee
Using the MSSM minimization conditions in \eqref{EWBC},
we can simplify the e.o.m.,
\be
\psi_1^0= {\mu v_2\over \sqrt{2}f}G+...\,,\ \psi_2^0= {\mu v_1\over \sqrt{2}f}G+...\,,\ \lambda_1=-{g_1(v_1^2-v_2^2)\over 4\sqrt{2}f}G+...\,,\ \lambda_2^3={g_2(v_1^2-v_2^2)\over 4\sqrt{2}f}G+...
\ee
After taking this into account, the Lagrangian of (\ref{Ldecaycomponents}) becomes,
\bea
\label{bla2}
\cL_{h\rightarrow \chi\,G}&=&-{1\over f} \big(m_1^2 G\psi_1h_1^{0*}+m_2^2 G\psi_2 h_2^{0*}\big)-{B\over f}\big( G\psi_2h_1^0+G\psi_1h_2^0 \big)\nonumber
\\
&&\!\!\!\!\!\!\!\!\!\!\!\!\!\!\!\!\!\!\!\!\!\!\!\!\!\!\!\!\!-{1\over f}\sum_{i=1,2}{M_i\over \sqrt{2}}\tilde{D}^a G \lambda^a_i
-{v_1^2-v_2^2\over 8f}g^2G(h_1^{0*}\psi_1^0-h_2^{0*}\psi_2^0)
-{\mu\over 2f}(g_2\lambda_2^3-g_1\lambda_1)G(h_1^{0*}v_2-h_2^{0*}v_1)
\nonumber\\
&&\!\!\!\!\!\!\!\!\!\!\!\!\!\!\!\!\!\!\!\!\!\!\!\!\!\!\!\!\!-{2c_1\over M^2} G \big( \partial^\mu h_1\partial_\mu \psi_2+\partial^\mu \psi_1\partial_\mu h_2 \big)+h.c.
\eea
In the second line of  \eqref{bla2} we see that, after electroweak symmetry breaking, additional terms contribute to the goldstino-Higgs-neutralino coupling, all of which are proportional to at least one power of $v$. 

In the third line of \eqref{bla2} we see terms contributing to the goldstino-Higgs-neutralino coupling, arising from the supersymmetric operator in the last line of \eqref{bla}, which are of higher order in derivatives. Hence, they correspond to higher-momentum corrections to the SUSY Goldberger-Treiman relation. 
By applying the e.o.m, we can extract the following leading order contribution to the goldstino-Higgs-neutralino coupling from the last term in \eqref{bla2},
\be
-{2c_1\over M^2}G\partial^\mu h_1\, \partial_\mu \psi_2=-{c_1\over M^2}\left(2 \mu^2 + m_1^2 \right) G h_1 \psi_2 + \cO(M^{-3}) \ . \label{gt}
\ee
One difference to the pion example is that, while the pion scenario only involves one fundamental scale, i.e.~$f_\pi$, in the SUSY example the suppression scale is not necessarily related to the SUSY breaking scale but rather to the scale of the massive states that have been integrated out in order to generate the SUSY operator in the last line of \eqref{bla}.
If $\mu \sim \msb$, we can again check, in agreement with all higher-dimensional operator effects discussed in our paper, that:\\
- in case i), the goldstino couplings in (\ref{gt}) are proportional to $\msb^4/f^2$, to be compared to the standard supercurrent couplings, proportional to $m_{soft}^2/f$. \\
 -  in case ii) with no dynamical sgoldstino and cutoff $\Lambda \sim \sqrt{f}$, the goldstino couplings in (\ref{gt}) are larger, since they are proportional to $(\msb/\sqrt{f})^3$. \\
Due to the structure of the coupling in (\ref{gt}), it provides a correction to the goldstino supercurrent couplings involving the $B$-term, i.e.~the $B$-operator in \eqref{Lgg1}, but it does not contribute to the $B$ soft term. 
Since it is of higher order in terms of the ratio of the soft parameter over $f$, it only becomes important when the soft parameter of the leading order term, which in this case is the $B$-parameter, is small. \\
For other aspects related to the analogy with pions, see e.g. \cite{luty}.

\section{Conclusions}

In this paper we studied various theoretical and phenomenological aspects and implications of low scale SUSY breaking scenarios.

We provided an extensive discussion concerning the coupling of the SUSY breaking sector to the MSSM, where we identified two different frameworks. In the first framework, SUSY breaking is mediated to the visible sector via a SUSY messenger sector with a characteristic energy scale $M$. The resulting effective Lagrangian has manifest SUSY, with SUSY being linearly realized both in the MSSM superfields, and in the goldstino supermultiplet, but spontaneously broken at the scale $f$ of the VEV of the auxiliary field  in the goldstino superfield. In the second framework, the visible sector couples to the SUSY breaking sector via a goldstino superfield and SUSY is realized non-linearly by using a SUSY constraint. The only scale in the minimal Lagrangian is $f$, while higher-dimensional operators, present in the general Lagrangian, are suppressed by a cut-off scale $\Lambda$, which can be generically as low as $\sqrt{f}$. We have shown that the higher-dimensional operators in the two frameworks provide corrections to the MSSM couplings, suppressed by the dimensionless ratio $(\msb/\sqrt{f})^n$, with the second framework delivering, in general, larger corrections than the first. The suppression by $(\msb/\sqrt{f})$ ensures the validity of the effective operator expansion.

We have also presented a systematic derivation of the most general form of the MSSM Lagrangian coupled to a goldstino supermultiplet, up to dimension-six operators. This was obtained after identifying and eliminating redundant operators by using field redefinitions.

The resulting irreducible Lagrangian was used in order to study the phenomenological implications of scenarios with a low SUSY breaking scale, in a model independent way. In particular, we concluded that the induced deviations of the Higgs couplings to gauge bosons from their SM values, can allow for a significant enhancement in the $h\to\gamma\gamma$ channel, while a smaller, but also considerable, deviation in the $h\to Z\gamma$ channel. The Higgs couplings to fermions are also modified by the presence of non-analytic, or ``wrong", Higgs Yukawa couplings. Their flavor structure is generally constrained by experimental bounds arising from flavor mixing processes. Such operators can affect the relation between the fermion masses and the corresponding Yukawa couplings.

The same superfield operators that deliver ``wrong" Higgs Yukawa couplings also contain one-goldstino couplings that contribute to monophoton+$\MET$ signatures. We compared the significance and the range of validity of this operator with respect to other effective operators that parametrize goldstino couplings with monophoton+$\MET$ signals, showing that the contribution from the ``wrong" Higgs Yukawa operator is most relevant for a neutralino that is dominantly higgsino. In contrast to this case, when the transverse momentum of the photon is low, below around 70 GeV, the dominant SUSY contribution to this final state arises from a Higgs boson decay into a goldstino and a neutralino, which subsequently decays into another goldstino and a photon.

Phenomenological bounds were also applied on the four-fermion contact interactions also included in the irreducible Lagrangian. Such interactions arise either from dimension-six operators or from lower-dimensional operators, upon integrating out the Higgs auxiliary fields, or the sgoldstino. These bounds have been published by the CMS and ATLAS collaborations on first-generation left-handed quark interactions, with interactions between two quarks and two leptons as well as on $uutt$-interactions. We demonstrated how such bounds are applicable to the ratio of the coefficients of effective operators and the cut-off scale.

The extension of the non-linear MSSM Lagrangian, in terms of higher-dimensional effective operators, provides corrections to the usual simple relations between the soft parameters, the SUSY breaking scale and the goldstino interaction couplings. This is in complete analogy with pion-nucleon interactions, in the context of chiral symmetry breaking, where it is well known that the most general effective Lagrangian, invariant under the chiral symmetry, should be considered before drawing any conclusions concerning the couplings. We exemplified this discussion by studying corrections to the goldstino-Higgs-neutralino coupling.

The work presented here consists of a model independent analysis of scenarios with a SUSY breaking scale in the range of a few TeV. It would be interesting to explore the possibility of constructing explicit SUSY breaking models that would allow for such a low SUSY breaking scale. Our analysis has shown that one of the most characteristic signatures of low scale SUSY breaking scenarios is the final state with a single photon and missing transverse energy. We argued that both an LHC search for this final state in the phase space regime where the transverse momentum of the photon is low (between around 45 GeV and 70 GeV, see \cite{Petersson:2012dp}), as well as an update on the LHC searches for this final state in the case where the photon momentum is high (around 100 GeV or higher), would be very important in terms of discovering/constraining low scale SUSY breaking scenarios.

\section*{Appendix}

\def\theequation{A-\arabic{equation}}
\def\thesubsection{A}
\setcounter{equation}{0}

\subsection{Details on the reduction of the low-scale effective action}
\label{appendixredefs}

Here we present the explicit results regarding the field redefinitions applied in section \ref{EffectiveActionLowScale}. The redefined parameters of the effective Lagrangian (\ref{L_TOT2}), after performing field redefinitions (\ref{derivredefs}) and (\ref{dimenredefs}), are
\be
\tilde{\zeta}_{i,I}=\zeta_{i,I}+\epsilon_{i,I}\,,\ \tilde{\zeta}'_{i,I}=\zeta'_{i,I}+|\epsilon_{i,I}|^2+\zeta^*_{i,I}\epsilon_{i,I}+\zeta_{i,I}\epsilon^*_{i,I}\,,\ \tilde{\zeta}''_{i,I}=\zeta''_{i,I}+\epsilon'_{i,I}+\epsilon_{i,I}\zeta_{i,I}\,,\ \tilde{\zeta}'''_{i,I}=\zeta'''_{i,I}+\delta^{'''*}_{i,I}\nn
\ee
\be
\tilde{a}=a+\epsilon_X-4{\hat{\mu}\over M}(\delta'''_1+\delta'''_2)\,,\quad\tilde{a}'=a'+\epsilon''_X+a(\epsilon_1+\epsilon_2)-4f_h(\delta'''_1+\delta'''_2)\,,\quad\tilde{a}''=a''\nn
\ee
\be
\tilde{s}_{0,u}=s_{0,u}+\epsilon_1^u\,,\ \tilde{s}_{u}=s_{u}-4\hat{y}_u\delta'_2+\epsilon_1^{u'}+\zeta_1\epsilon_1^u+s_{0,u}(\epsilon_u+\epsilon_q)\,,\ \tilde{s}'_{u}=s'_{u}-4\hat{y}_u\delta''_2+s_{0,u}\epsilon_1^*+\epsilon_1^*\epsilon_1^u+\zeta_1^*\epsilon_1^u\nn
\ee
\be
\tilde{s}_{0,d}=s_{0,d}+\epsilon_2^d\,,\ \tilde{s}_{d}=s_{d}+4\hat{y}_d\delta'_1+\epsilon_2^{d'}+\zeta_2\epsilon_2^d+s_{0,d}(\epsilon_d+\epsilon_q)\,,\ \tilde{s}'_{d}=s'_{d}+4\hat{y}_d\delta''_1+s_{0,d}\epsilon_2^*+\epsilon_2^*\epsilon_2^d+\zeta_2^*\epsilon_2^d\nn
\ee
\be\label{tildes1}
\tilde{s}_{0,e}=s_{0,e}+\epsilon_2^e\,,\ \tilde{s}_{e}=s_{e}+4\hat{y}_e\delta'_1+\epsilon_2^{e'}+\zeta_2\epsilon_2^e+s_{0,e}(\epsilon_e+\epsilon_l)\,,\ \tilde{s}'_{e}=s'_{e}+4\hat{y}_e\delta''_1+s_{0,e}\epsilon_2^*+\epsilon_2^*\epsilon_2^e+\zeta_2^*\epsilon_2^e
\ee
\be
\tilde{c}_1=c_1+2\delta^{'''*}_X\,,\quad\tilde{c}_2=c_2+\delta^{''*}_2+\delta^{'''*}_X-\delta_1^*\epsilon_1+\delta_1^*\zeta_1\nn
\ee
\be\label{tildes2}
\tilde{c}_{0,3}=c_{0,3}-\delta_1^*+\delta_2^*\,,\quad\!\!\! \tilde{c}_3=c_3+\delta^{''*}_1+\delta^{'''*}_X+c_{0,3}\epsilon_2+\zeta_2\delta_2^*+\epsilon_2\delta_2^*\,,\quad\!\!\!\tilde{c}'_3=c'_3-\delta^{'*}_1+\delta^{'*}_2-\delta_1^*\zeta_1^*+\delta_2^*\zeta_2^*
\ee

\bea\label{tildes3}
&&\tilde{\hat{\mu}}=\hat{\mu}+{f\over M}\epsilon_X\,,\quad\tilde{f}_h=f_h+(\epsilon_1+\epsilon_2){\hat{\mu}\over M}+\epsilon''_X{f\over M^2}
\\
&&\tilde{f}'_h\!=\!f'_h\!+\!f_h(\epsilon_1\!+\!\epsilon_2)\!+\!(\epsilon'_1\!+\!\epsilon'_2){\hat{\mu}\over M}\!+\!\epsilon_1\epsilon_2{\hat{\mu}\over M}\,,\quad\!\!\tilde{f}''_h\!=\!f''_h\!+\!f'_h(\epsilon_1\!+\!\epsilon_2)\!+\!f_h(\epsilon'_1\!+\!\epsilon'_2)+f_h\epsilon_1\epsilon_2\nn
\eea
\bea\label{tildes4}
&&\tilde{\hat{y}}_u=\hat{y}_u+{f\over M^2}\epsilon^u_X-{\hat{\mu}\over M}\epsilon_1^u\nn
\\
&&\tilde{f}_u=f_u+(\epsilon_2+\epsilon_q+\epsilon_u)\hat{y}_u-\epsilon^{u'}_1{\hat{\mu}\over M}-f_h\epsilon_1^u-{\hat{\mu}\over M}\epsilon_2\epsilon_1^u\nn
\\
&&\tilde{f}'_u=f'_u\!+\!(\epsilon_2\!+\!\epsilon_q\!+\!\epsilon_u)f_u\!+\!(\epsilon'_2\!+\!\epsilon'_q\!+\!\epsilon'_u+\epsilon_2\epsilon_q+\epsilon_2\epsilon_u+\epsilon_q\epsilon_u)\hat{y}_u-f_h\epsilon^{u'}_1-f'_h\epsilon_1^u-f_h\epsilon_2\epsilon_1^u\nn
\\
&&\tilde{f}''_u=f''_u+\hat{y}_u(\delta'''_2+\delta'''_q+\delta'''_u)+{1\over 4}a\,\epsilon_1^u-{1\over 4}\epsilon_X^u
\eea
\bea\label{tildes5}
&&\tilde{\hat{y}}_d=\hat{y}_d+{f\over M^2}\epsilon^d_X-{\hat{\mu}\over M}\epsilon_2^d\nn
\\
&&\tilde{f}_d=f_d+(\epsilon_1+\epsilon_q+\epsilon_d)\hat{y}_d-\epsilon^{d'}_2{\hat{\mu}\over M}-f_h\epsilon_2^d-{\hat{\mu}\over M}\epsilon_1\epsilon_2^d\nn
\\
&&\tilde{f}'_d\!=\!f'_d\!+\!(\epsilon_1\!+\!\epsilon_q\!+\!\epsilon_d)f_d\!+\!(\epsilon'_1\!+\!\epsilon'_q\!+\!\epsilon'_d\!+\!\epsilon_1\epsilon_q\!+\!\epsilon_1\epsilon_d\!+\!\epsilon_q\epsilon_d)\hat{y}_d\!-\!f_h\epsilon^{d'}_2\!-\!f'_h\epsilon_2^d-f_h\epsilon_1\epsilon_2^d\nn
\\
&&\tilde{f}''_d=f''_d+\hat{y}_d(\delta'''_1+\delta'''_q+\delta'''_d)+{1\over 4}a\,\epsilon_2^d-{1\over 4}\epsilon_X^d
\eea
\bea\label{tildes6}
&&\tilde{\hat{y}}_e=\hat{y}_e+{f\over M^2}\epsilon^e_X-{\hat{\mu}\over M}\epsilon_2^e\nn
\\
&&\tilde{f}_e=f_e+(\epsilon_1+\epsilon_l+\epsilon_e)\hat{y}_e-\epsilon^{e'}_2{\hat{\mu}\over M}-f_h\epsilon_2^e-{\hat{\mu}\over M}\epsilon_1\epsilon_2^e\nn
\\
&&\tilde{f}'_e\!=\!f'_e\!+\!(\epsilon_1\!+\!\epsilon_l\!+\!\epsilon_e)f_e\!+\!(\epsilon'_1\!+\!\epsilon'_l\!+\!\epsilon'_e\!+\!\epsilon_1\epsilon_l\!+\!\epsilon_1\epsilon_e\!+\!\epsilon_l\epsilon_e)\hat{y}_e\!-\!f_h\epsilon^{e'}_2-f'_h\epsilon_2^e-f_h\epsilon_1\epsilon_2^e\nn
\\
&&\tilde{f}''_e=f''_e+\hat{y}_e(\delta'''_1+\delta'''_l+\delta'''_e)+{1\over 4}a\,\epsilon_2^e-{1\over 4}\epsilon_X^e
\eea
\bea\label{tildes7}
&&\tilde{\hat{y}}_w^a=\hat{y}_w^a+{f\over M^2}\epsilon^{w\,a}_X\,,\quad\tilde{f}_w^a=f_w^a\,,\quad\tilde{f}^{a\,'}_w=f^{a\,'}_w\,,\quad\tilde{f}^{a\,''}_w=f^{a\,''}_w-{1\over 4}\epsilon_X^{w\,a}
\eea
\bea\label{tildes8}
&&\hspace{-0.7cm}\tilde{t}_{0,1}=t_{0,1}+f_h\epsilon_X+{\hat{\mu}\over M}(\epsilon''_1+\epsilon''_2)
\\
&&\hspace{-0.7cm}\tilde{t}_{1}=t_{1}+2t_{0,1}(\epsilon_1+\epsilon_2)+f_h(\epsilon''_1+\epsilon''_2)+2f'_h\epsilon_X+f_h\epsilon_1\epsilon_X+f_h\epsilon_2\epsilon_X+f_h\epsilon''_X\nn
\\
&&\hspace{-0.7cm}\tilde{t}_{0,2}=t_{0,2}+\hat{y}_u\epsilon_2^d+\hat{y}_d\epsilon_1^u-{\hat{\mu}\over M}\epsilon_1^u\epsilon_2^d\nn
\\
&&\hspace{-0.7cm}\tilde{t}_{2}\!=\!t_{2}\!+\!t_{0,2}(2\epsilon_q\!+\!\epsilon_d\!+\!\epsilon_u)\!+\!\hat{y}_u(\epsilon_2^{d'}\!+\!\epsilon_2^d\epsilon_q\!+\!\epsilon_2^d\epsilon_u)+\hat{y}_d(\epsilon_1^{u'}\!+\!\epsilon_1^u\epsilon_q\!+\!\epsilon_1^u\epsilon_d)\!+\!f_u\epsilon_2^d\!+\!f_d\epsilon_1^u\!-\!f_h\epsilon_2^d\epsilon_1^u\nn
\\
&&\hspace{-0.7cm}\tilde{t}_{0,3}=t_{0,3}+\hat{y}_u\epsilon_2^e+\hat{y}_e\epsilon_1^u-{\hat{\mu}\over M}\epsilon_1^u\epsilon_2^e\nn
\\
&&\hspace{-0.7cm}\!\tilde{t}_{3}\!=\!t_{3}\!+\!t_{0,3}(\epsilon_q\!+\!\epsilon_u\!+\!\epsilon_l\!+\!\epsilon_e)\!+\!\hat{y}_u(\epsilon_2^{e'}\!+\!\epsilon_2^e\epsilon_q\!+\!\epsilon_2^e\epsilon_u)\!+\!\hat{y}_e(\epsilon_1^{u'}\!+\!\epsilon_1^u\epsilon_e\!+\!\epsilon_1^u\epsilon_l)\!+\!f_u\epsilon_2^e\!+\!f_e\epsilon_1^u\!-\!f_h\epsilon_2^e\epsilon_1^u\nn
\eea

Many of the higher dimensional operators are eliminated by appropriately fixing the redefinition parameters. The exact relations for the choice that we have made are
\bea\label{elimrelations}
&&\hspace{-0.7cm}\delta_1^*-\delta_2^*=c_{0,3}\,,\ \delta_1^{'*}=\delta_2^{'*}+c_3'-\delta_1^*\zeta_1^*+\delta_2^*\zeta_2^*\,,\ \delta_2'\,\hat{y}_u={1\over 4}\left( s_u+\epsilon_1^{u'}+\zeta_1\epsilon_1^u+s_{0,u}(\epsilon_u+\epsilon_q) \right)\nn
\\
&&\hspace{-0.7cm}\delta_1^{''*}=-\delta_X^{'''*}-c_{0,3}\epsilon_2-c_3-\delta_2^*(\zeta_2+\epsilon_2)\,,\quad \delta_2^{''*}=-c_2-\delta_X^{'''*}+\delta_1^*(\epsilon_1-\zeta_1)\nn
\\
&&\hspace{-0.7cm}\delta^{'''*}_{i,I}=-\zeta'''_{i,I}\,,\quad\!\! \delta_X^{'''*}=-{c_1\over 2}\nn
\\
&&\hspace{-0.7cm}\epsilon_{i,I}=-\zeta_{i,I}\,,\quad\!\! \epsilon'_{i,I}=-\zeta''_{i,I}-\epsilon_{i,I}\zeta_{i,I}\nn
\\
&&\hspace{-0.7cm}\epsilon_2^{d'}=-\left( s_d+4\hat{y}_d\delta_1'+\zeta_2\epsilon_2^d+s_{0,d}(\epsilon_d+\epsilon_q)\right)\nn
\\
&&\hspace{-0.7cm}\epsilon_2^{e'}=-\left( s_e+4\hat{y}_e\delta_1'+\zeta_2\epsilon_2^e+s_{0,e}(\epsilon_e+\epsilon_l)\right)\nn
\\
&&\hspace{-0.7cm}\epsilon_2^d\,\hat{y}_u=-\left( t_{0,2}+\hat{y}_d\epsilon_1^u \right)\,,\quad\!\! \epsilon_2^e\,\hat{y}_u=-\left( t_{0,3}+\hat{y}_e\epsilon_1^u \right)\nn
\\
&&\hspace{-0.7cm}\hat{y}_d\epsilon_1^{u'}\!=\!-\!\left[ t_2\!+\!t_{0,2}(2\epsilon_q\!+\!\epsilon_d\!+\!\epsilon_u)\!+\!\hat{y}_u(\epsilon_2^{d'}\!+\!\epsilon_2^d(\epsilon_q\!+\!\epsilon_u))\!+\!\hat{y}_d\,\epsilon_1^u(\epsilon_q\!+\!\epsilon_d)\!+\!f_u\epsilon_2^d\!+\!f_d\epsilon_1^u\!-\!f_h\epsilon_2^d\epsilon_1^u \right]\nn
\\
&&\hspace{-0.7cm}\hat{y}_e\epsilon_1^u\epsilon_{e}\!\!=\!-\!\!\left[ t_3\!+\!t_{0,3}(\!\epsilon_q\!+\!\epsilon_u\!+\!\epsilon_l\!+\!\epsilon_e)\!+\!\hat{y}_u\!(\epsilon_2^{e'}\!+\!\epsilon_2^e(\epsilon_q\!+\!\epsilon_u))\!+\!\hat{y}_e(\epsilon_1^{u'}\!+\!\epsilon_1^u\epsilon_l)\!+\!f_u\epsilon_2^e\!+\!f_e\epsilon_1^u\!-\!f_h\epsilon_2^e\epsilon_1^u\!  \right]\nn
\\
&&\hspace{-0.7cm}\epsilon_X^e=4(f''_e+\hat{y}_e(\delta_1'''+\delta_e'''+\delta_l''')+{1\over 4}a\,\epsilon_2^e)\,,\quad\epsilon_X^d=4(f''_d+\hat{y}_d(\delta_1'''+\delta_d'''+\delta_q''')+{1\over 4}a\,\epsilon_2^d)\nn
\\
&&\hspace{-0.7cm}\epsilon_X^u=4(f''_u+\hat{y}_u(\delta_2'''+\delta_q'''+\delta_u''')+{1\over 4}a\,\epsilon_1^u)\,,\quad \epsilon_X^{w\,a}=4f^{a''}_w\nn
\\
&&\hspace{-0.7cm}\epsilon_X=-a\,,\quad \epsilon_X''=-a'-a(\epsilon_1+\epsilon_2)+4f_h(\delta_1'''+\delta'''_2)
\eea

\def\theequation{B-\arabic{equation}}
\def\thesubsection{B}
\setcounter{equation}{0}

\subsection{The set of dimension-six operators}
\label{appendixL6}

The full set of dimension-6 operators $\cL_6$ that do not involve the goldstino multiplet is given by
\be
\cL_6=\int d^4\theta\ \cD_6+\left( \int d^2\theta\ \cF_6+h.c.\right)
\ee
where
\be
M^2 \cF_6=(f_6^uH_2 Q U^c+f_6^dQD^cH_1+f_6^eLE^cH_1+f_6^{w\,a}W^aW^a)H_1H_2
\ee
and
\bea
&&\hspace{-1.2cm}M^2 \cD_6=H_i^\dag e^{V_i}H_i\left(d_6^{ij}H_j^\dag e^{V_j}H_j+(d_6^iH_1H_2+h.c.)+d_6^{iI}\Phi_I^\dag e^{V_I}\Phi_I\right)+d_6^h|H_1H_2|^2\nn
\\
&&\hspace{-1.2cm}+(d_6(LE^c)^\dag e^{V_2}QD^c+h.c.)+d_6'(QU^c)^\dag e^{V_1}QU^c+d_6''(QD^c)^\dag e^{V_2}QD^c+d_6'''(LE^c)^\dag e^{V_2}LE^c\nn
\\
&&\hspace{-1.2cm}+\ \Phi_I^\dag e^{V_I} \Phi_I\left(d_6^{IJ}\Phi_J^\dag e^{V_J} \Phi_J+(d_6^IH_1H_2+h.c.)\right)\,.
\eea
The coefficients are in principle of order 1, while all flavor and gauge indices are suppressed. More details on MSSM phenomenology with higher-dimensional operators can be found in \cite{Espinoza,dst,adgt}.

\section*{Acknowledgements}
We thank Ignatios Antoniadis, Lorenzo Calibbi, Gero von Gersdorff, Dumitru Ghilencea, Carlos Pena, Alberto Romagnoni and Riccardo Torre for useful discussions. The work presented was supported in part by the European ERC Advanced Grant 226371 MassTeV, by the grants ANR-05-BLAN-0079-02 and the PITN contract PITN-GA-2009-237920. The work of C.P. is supported in part by IISN-Belgium (conventions 4.4511.06, 4.4505.86 and 4.4514.08), by the ``Communaut\'e Fran\c{c}aise de Belgique" through the ARC program and by a ``Mandat d'Impulsion Scientifique" of the F.R.S.-FNRS.

\vspace*{0.7cm}


\end{document}